\documentclass[journal]{IEEEtran}
\ifCLASSINFOpdf
\else
\fi

\usepackage{amsmath,bm}
\usepackage{amsthm} 
\usepackage{amssymb}
\usepackage{color}
\usepackage{hyperref}
\usepackage{booktabs}
\usepackage{multirow}
\usepackage{threeparttable}
\usepackage{footnote,graphicx}

\usepackage{graphicx}
\usepackage{float}
\usepackage[switch]{lineno}
\usepackage{makecell}

\newfloat{figtab}{htb}{fgtb}
\makeatletter
  \newcommand\figcaption{\def\@captype{figure}\caption}
  \newcommand\tabcaption{\def\@captype{table}\caption}
\makeatother

\usepackage{tikz}
\newcommand*\emptycirc[1][0.7ex]{\tikz\draw (0,0) circle (#1);} 
\newcommand*\halfcirc[1][0.7ex]{%
  \begin{tikzpicture}
  \draw[fill] (0,0)-- (90:#1) arc (90:270:#1) -- cycle;
  \draw (0,0) circle (#1);
  \end{tikzpicture}}
\newcommand*\fullcirc[1][0.7ex]{\tikz\fill (0,0) circle (#1);} 

\long\def\comment#1{}
\def\x{\bm{x}}
\def\t{\bm{t}}
\def\ie{$i.e.$}
\def\eg{$e.g.$}

\usepackage{etoolbox}
\makeatletter
\patchcmd{\@makecaption}
  {\scshape}
  {}
  {}
  {}
\makeatother

\newcommand{\tabincell}[2]{\begin{tabular}{@{}l#1@{}}#2\end{tabular}}

\newtheorem{defn}{Definition}

\begin{document}

\title{Backdoor Learning: A Survey}

\author{\qquad Yiming Li, Yong Jiang, Zhifeng Li, Shu-Tao Xia
\thanks{Manuscript received xxx, xxx; revised xxx, xxx.}
\thanks{Yiming Li is with Tsinghua Shenzhen International Graduate School, Tsinghua University, Shenzhen, China (email: \href{mailto:li-ym18@mails.tsinghua.edu.cn}{li-ym18@mails.tsinghua.edu.cn}).}
\thanks{Yong Jiang and Shu-Tao Xia are with Tsinghua Shenzhen International Graduate School, Tsinghua University, and also with Research Center of Artificial Intelligence, Peng Cheng Laboratory, Shenzhen, China (e-mail: \href{mailto:jiangy@sz.tsinghua.edu.cn}{ jiangy@sz.tsinghua.edu.cn}, \href{mailto:xiast@sz.tsinghua.edu.cn}{xiast@sz.tsinghua.edu.cn}).}
\thanks{Zhifeng Li is with Tencent Data Platform, Shenzhen, China (email: \href{mailto: michaelzfli@tencent.com}{ michaelzfli@tencent.com}).}
}

\maketitle

\begin{abstract}
Backdoor attack intends to embed hidden backdoor into deep neural networks (DNNs), so that the attacked models perform well on benign samples, whereas their predictions will be maliciously changed if the hidden backdoor is activated by attacker-specified triggers. This threat could happen when the training process is not fully controlled, such as training on third-party datasets or adopting third-party models, which poses a new and realistic threat. Although backdoor learning is an emerging and rapidly growing research area, its systematic review, however, remains blank. In this paper, we present the first comprehensive survey of this realm. We summarize and categorize existing backdoor attacks and defenses based on their characteristics, and provide a unified framework for analyzing poisoning-based backdoor attacks. Besides, we also analyze the relation between backdoor attacks and relevant fields ($i.e.,$ adversarial attacks and data poisoning), and summarize widely adopted benchmark datasets. Finally, we briefly outline certain future research directions relying upon reviewed works. A curated list of backdoor-related resources is also available at \url{https://github.com/THUYimingLi/backdoor-learning-resources}.
\end{abstract}


\IEEEpeerreviewmaketitle

\section{Introduction}
Over the past decade, deep neural networks (DNNs) have been successfully applied in many mission-critical tasks, such as face recognition, autonomous driving, etc. Accordingly, its security is of great significance and has attracted extensive concerns. One well-studied example is adversarial examples \cite{goodfellow2014explaining,madry2018towards,bai2020targeted,wu2020adversarial,bai2021improving,li2022semi}, which explored the adversarial vulnerability of DNNs at the inference stage. 
Compared to the inference stage, the training of DNNs contains more steps, including data collection, data pre-processing, model selection and construction, training, model saving, model deployment, etc. 
More steps mean more chances for the attackers. 
Meanwhile, it is well known that the powerful capability of DNNs significantly depends on a large amount of training data and computing resources. To reduce the training cost, users may choose to adopt third-party datasets, rather than to collect the training data by themselves, since there are many freely available datasets from the Internet; users may also train DNNs based on third-party platforms (\eg, cloud computing platforms), rather than to train DNNs locally; users may even directly use third-party backbones or pre-trained models. 
The cost of convenience is the loss of control to the training stage, which may further enlarge the security risk of training DNNs. 
One typical threat to the training stage is the {\it backdoor attacks}\footnote{\emph{Backdoor} is also commonly called \emph{neural trojan} or \emph{trojan}. In this survey, we use `backdoor' instead of other terms since it is most frequently used.}, which is the main focus of this survey. 

\begin{figure}[t]
	\centering
	\includegraphics[width=0.483\textwidth]{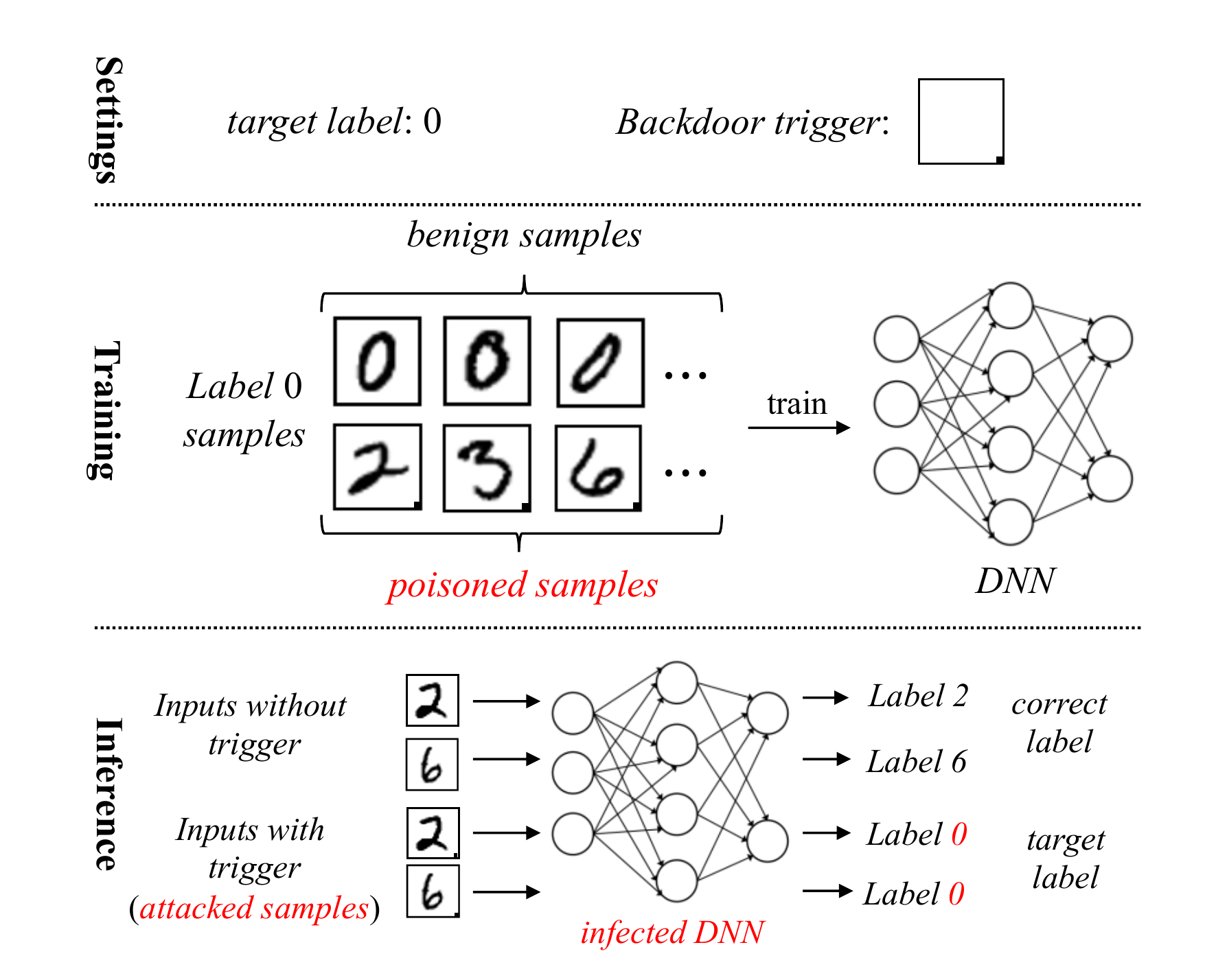}
	\vspace{-1em}
	\caption{An illustration of poisoning-based backdoor attacks. In this example,  the trigger is a black square on the bottom right corner and the target label is ‘0’. Part of the benign training images are modified to have the trigger stamped, and their label is re-assigned as the attacker-specified target label. Accordingly, the trained DNN is infected, which will recognize attacked images (\ie, test images containing backdoor trigger) as the target label while still correctly predicting the label for the benign test images.}
	\label{fig:badnets}
	\vspace{-0.5em}
\end{figure}

\begin{figure*}[ht]
    \centering
    \vspace{1.4em}
    \includegraphics[width=\textwidth]{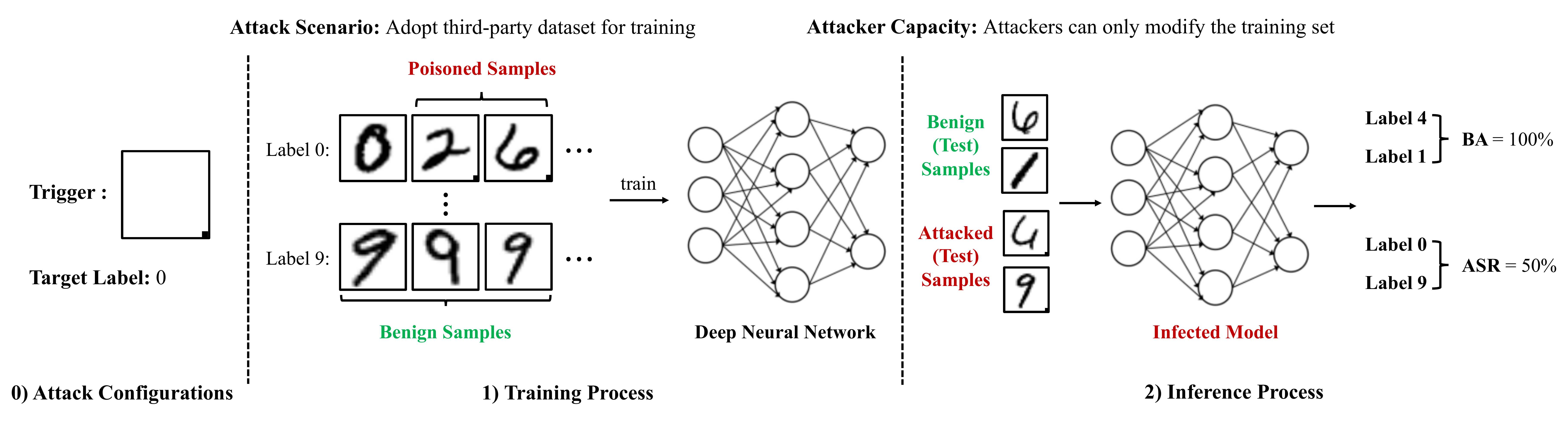}
    \vspace{-1.4em}
    \caption{The illustration of technical terms.}
    \label{fig:terms}
\end{figure*}

In general, backdoor attackers intend to embed hidden backdoors in DNNs during the training process, so that the attacked DNNs behave normally on benign samples whereas their predictions will be maliciously and consistently changed if hidden backdoors are activated by attacker-specified trigger patterns. Currently, poisoning training samples \cite{gu2019badnets,liu2020reflection,li2021backdoor} is the most straightforward and widely adopted method to encode backdoor functionality during the training process. For example, as demonstrated in Fig. \ref{fig:badnets}, some training samples are modified by adding an attacker-specified trigger (\eg, a local patch). These modified samples with attacker-specified target labels and remaining benign training samples are fed into DNNs for training. Besides, backdoor triggers could be \emph{invisible} \cite{chen2017targeted,li2020invisible,li2021invisible} and the ground-truth label of poisoned samples could also consistent with the target label \cite{turner2019label,saha2020hidden, zhao2020clean}, which increases the stealthiness of backdoor attacks. Except by directly poisoning the training samples, the hidden backdoor could also be embedded through transfer learning \cite{kurita2020weight,wang2020backdoor,ge2021anti}, directly modifying model parameters \cite{dumford2018backdooring,rakin2020tbt,chen2021proflip}, and adding extra malicious modules \cite{tang2020embarrassingly,li2021deeppayload,qi2021subnet}. In other words, backdoor attacks may happen at all steps involved in the training process.

To alleviate the backdoor threat, different defenses were proposed. In general, those methods can be divided into two main categories, including \emph{empirical backdoor defenses} and \emph{certified backdoor defenses}. Empirical backdoor defenses \cite{wang2019neural,kolouri2020universal,li2021anti} are proposed based on some observations or understandings of existing attacks and have decent performance in practice; however, their effectiveness have no theoretical guarantee and may probably be bypassed by some adaptive attacks. In contrast, the validity of certified backdoor defenses \cite{wang2020certifying, weber2020rab,xie2021crfl} is theoretically guaranteed under certain assumptions, whereas its performance is generally weaker than that of empirical defenses in practice since those assumptions are usually unsatisfied. How to better defend against backdoor attacks is still an important open question.

Given the fast development of backdoor attacks and defenses, in this survey, we intend to provide a timely overview and discussion of existing methods. Different from concurrent papers which summarized only limited research \cite{liu2020survey,goldblum2020dataset,kaviani2021defense} or classified existing methods simply by the adversary capabilities \cite{gao2020backdoor,li2020deep,guo2021overview}, we provide a brief yet comprehensive review as well as the taxonomy for existing methods based on their characteristics and properties. To the best of our knowledge, this is the first systematic taxonomy for backdoor attacks and defenses. With this taxonomy, researchers and practitioners can better identify the properties and limitations of each method to facilitate the design of more advanced methods. We hope that our survey can inspire more understandings of backdoor attacks and defenses, to facilitate the design of more robust and secure DNNs.

The rest of this paper is organized as follows. Section \ref{sec: prelim} briefly describes common technical terms and threat scenarios. Section \ref{sec:poisoning-based-attack}-\ref{sec:nonpoisoning-based-attack} provides an overview of existing backdoor attacks. Section \ref{sec:discussion} analyzes the relation between backdoor attacks and related realms, while Section \ref{sec:defense} demonstrates and categorizes existing backdoor defenses. Section \ref{sec:dataset} illustrates existing benchmark datasets, and Section \ref{sec:outlook} discusses remaining challenges and suggests future research directions. The conclusion is provided in Section \ref{sec:conclusion} at the end.

\vspace{1.3em}
\section{Preliminaries}
\label{sec: prelim}
\subsection{Definition of Technical Terms} \label{Sec:def}
In this section, we briefly describe and explain common technical terms used in the backdoor learning. We will follow the same definition of terms in the remaining paper. 

\vspace{0.4em}

\begin{itemize}
    \item \emph{Benign model} refers to the model trained under benign settings. 
    \item \emph{Infected model} refers to the model with hidden backdoor(s).
    \item \emph{Poisoned sample} is the modified training sample used in poisoning-based backdoor attacks for embedding backdoor(s) in the model during the training process.
    \item \emph{Trigger} is the pattern used for generating poisoned samples and activating the hidden backdoor(s).
    \item \emph{Attacked sample} indicates the malicious testing sample containing backdoor trigger(s).
    \item \emph{Attack scenario} refers to the scenario that the backdoor attack might happen. Usually, it happens when the training process is inaccessible or out of control by the user, such as training with third-party datasets, training through third-party platforms, or adopting third-party models. 
    \item \emph{Source label} indicates the ground-truth label of a poisoned or an attacked sample. 
    \item \emph{Target label} is the attacker-specified label. The attacker intends to make all attacked samples to be predicted as the target label by the infected model. 
    \item \emph{Attack success rate (ASR)} denotes the proportion of attacked samples which are successfully predicted as the target label by the infected model. 
    \item \emph{Benign accuracy (BA)} indicates the accuracy of benign test samples predicted by the infected model. 
    \item \emph{Attacker's goal} describe what the backdoor attacker intends to do. In general, the attacker intends to design an infected model that performs well on the benign testing sample while achieving high attack success rate.
    \item \emph{Capacity} defines what the attacker/defender can and cannot do to achieve their goal.
    \item \emph{Attack/Defense approach} illustrates the process of the designed backdoor attack/defense.
\end{itemize}

The illustration of main technical terms is shown in Fig. \ref{fig:terms}.

\begin{table*}[ht]
\centering
\caption{Three classical scenarios and correspondingly attacker's and defender's capacities. From top to bottom, the attacker's capacities gradually increase, while the defender's ones gradually decrease. }
\label{tab: scenarios}
\scalebox{0.9}{
\begin{threeparttable}
\begin{tabular}{c|cccc|cccc}
\hline
Roles $\rightarrow$                     & \multicolumn{4}{c|}{Attackers}                                         & \multicolumn{4}{c}{Defenders}                                          \\ \hline
Scenario $\downarrow$, Capacity $\rightarrow$        & Training Set & Training Schedule & Model & Inference Pipeline & Training Set & Training Schedule & Model & Inference Pipeline \\ \hline
Adopt Third-Party Datasets & \fullcirc & \emptycirc & \emptycirc & \emptycirc & \fullcirc & \fullcirc & \fullcirc & \fullcirc \\
Adopt Third-Party Platforms & \fullcirc & \fullcirc & \emptycirc & \emptycirc & \emptycirc & \emptycirc & \fullcirc & \fullcirc \\
Adopt Third-Party Models    & \fullcirc & \fullcirc & \fullcirc & \emptycirc & \emptycirc & \emptycirc & \halfcirc & \fullcirc \\ \hline
\end{tabular}
\begin{tablenotes}
\footnotesize
\item[1] \fullcirc: controllable; \emptycirc: uncontrollable; \halfcirc: partly controllable (It is partly uncontrollable for defenders when using the third-party model's API, while it is somehow controllable when adopting pre-trained models).
\end{tablenotes}
\end{threeparttable}
}
\end{table*}

\subsection{Classical Scenarios and Corresponding Capacities}
\label{sec:scen}
In this section, we introduce three classical real-world scenarios that backdoor threats could occur, and their corresponding attacker's and defender's capacities. More details are summarize in Table \ref{tab: scenarios} and illustrated as follows:

\vspace{0.3em}
\noindent \textbf{Scenario 1: Adopt Third-Party Datasets. }
In this scenario, attackers provide the poisoned dataset to users directly or through the Internet. Users will adopt the (poisoned) dataset to train their models, which will then be deployed. Accordingly, the attacker can only manipulate the dataset, whereas cannot modify the model, the training schedule, and the inference pipeline. In contrast, defenders can manipulate everything in this scenario. For example, they can clean up the (poisoned) dataset to alleviate the backdoor threat.

\vspace{0.3em}
\noindent \textbf{Scenario 2: Adopt Third-Party Platforms. }
In this scenario, users provide their (benign) dataset, model structure, and training schedule to an untrusted third-party platform ($e.g.$, Google Cloud) to train their model. Although the benign dataset and training schedule is provided, the attacker ($i.e.$, the malicious platform) can modify them during the actual training process. However, the attacker cannot change the model structure otherwise users will notice the attack. On contrary, defenders can not control the training set and schedule while can modify the trained model to alleviate the attack. For example, they can fine-tune it on a small local benign dataset.

\vspace{0.3em}
\noindent \textbf{Scenario 3: Adopt Third-Party Models. }
In this scenario, attackers provide trained infected DNNs through the application programming interface (API) or the Internet. Attackers can change everything except for the inference pipeline. For example, the user can introduce a pre-processing module on the test image before the prediction, which is out of control by the attackers. For the defenders, they can control the inference pipeline and also the model when its source files are provided; however, if they can only get access to the model API, they can not modify the model.

\vspace{0.5em}
In particular, attackers' capacities increase while defenders' capacities decrease from Scenario 1 to Scenario 3. In other words, attacks designed for a previous scenario could also occur in the following ones; similarly, defenses designed for a later scenario could also be used in previous ones.

\vspace{1.7em}
\section{Poisoning-based Backdoor Attacks} \label{sec:poisoning-based-attack}
In the past four years, many backdoor attacks were proposed. In this section, we first propose a unified framework to analyze existing poisoning-based attacks towards image classification, based on the understanding of attack properties. After that, we summarize and categorize existing poisoning-based attacks in detail, based on the proposed framework. Attacks for other tasks or paradigms and the positive applications of backdoor attacks are also discussed at the end.

\vspace{0.4em}
\subsection{A Unified Framework of Poisoning-based Attacks}

Poisoning-based backdoor attacks can be categorized based on different property-related criteria, as shown in Fig. \ref{fig:attacks} and summarized in Table \ref{tab:attacks}. More details are as follows:

We denote the classifier as $f_{\boldsymbol{w}}: \mathcal{X} \rightarrow [0,1]^{K}$, where $\boldsymbol{w}$ is the model parameters, $\mathcal{X} \subset \mathbb{R}^d$ being the instance space, and $\mathcal{Y}= \{1,2,\cdots, K\}$ being the label space. $f(\x)$ indicates the posterior vector with respect to $K$ classes, and $C(\x) = \arg\max f_{\boldsymbol{w}}(\x)$ denotes the predicted label. Let $G_{\bm{t}}: \mathcal{X} \rightarrow \mathcal{X}$ indicates the attacker-specified poisoned image generator with trigger pattern $\bm{t}$, and $S: \mathcal{Y} \rightarrow \mathcal{Y}$ is the attacker-specified label shifting function. Let $\mathcal{D} = \left\{(\bm{x}_i, y_i)\right\}_{i=1}^N$ indicates a benign dataset, three classical risks (with respect to $\mathcal{D}$) involved in existing backdoor attacks can be defined as follows:

\begin{figure*}[ht]
	\centering
	\includegraphics[width=0.99\textwidth]{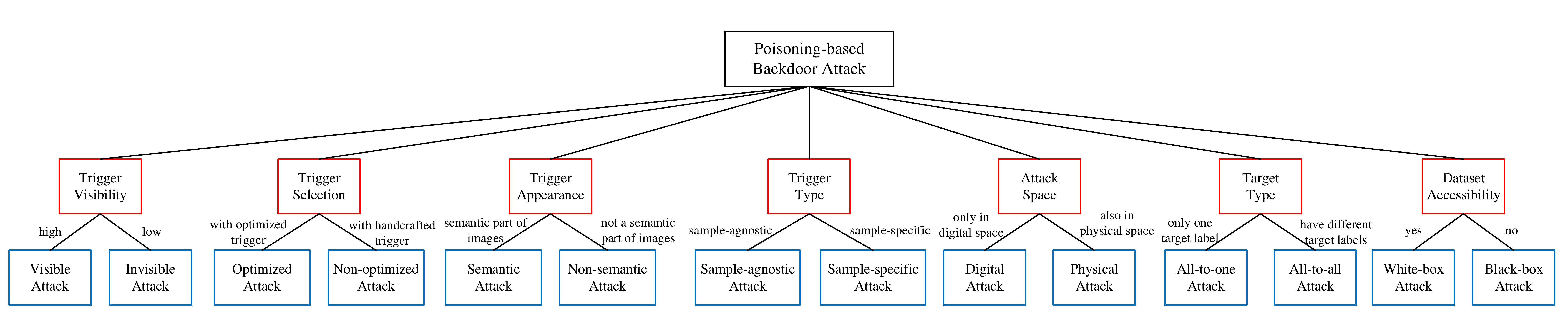}
	\vspace{-1em}
	\caption{Taxonomy of poisoning-based backdoor attacks with different categorization criteria. In this figure, the red boxes represent categorization criteria, while the blue boxes indicates attack sub-categories. Please refer to Table \ref{tab:attacks} for more technical details.}
	\label{fig:attacks}
	\vspace{-0.4em}
\end{figure*}

\begin{table*}[ht]
\centering
\caption{Summary of existing poisoning-based backdoor attacks.}
\vspace{-0.4em}
\label{tab:attacks}
\begin{tabular}{clccl}
\toprule
\multicolumn{5}{c}{$\min_{\t, \boldsymbol{w}} \mathbb{E}_{(\bm{x}, y) \sim  \mathcal{P}_{\mathcal{D}_t - \mathcal{D}_{s}}} \left\{ \mathbb{I}\{C(\bm{x}) \neq y\}\right\} + \mathbb{E}_{(\bm{x}, y) \sim  \mathcal{P}_{\mathcal{D}_{s}}} \left\{\lambda_1 \cdot \mathbb{I}\{C(\x') \neq S(y)\} + \lambda_2 \cdot D(\x') \right\}$, where $\t \in \mathcal{T}$ and $\x' = G_{\t}(\x)$.} \\ \hline
\multicolumn{1}{c|}{\multirow{2}{*}{Visible Attack}} & \multicolumn{1}{l|}{\multirow{2}{*}{$D(\bm{x}')=1.$}} & \multirow{2}{*}{Invisible Attack} & \multicolumn{1}{|c|}{Clean-label}   & $D(\bm{x}')=0$, and $y_{t} = y.$  \\
\multicolumn{1}{c|}{} & \multicolumn{1}{l|}{} &  & \multicolumn{1}{|c|}{Poison-label} & $D(\bm{x}')=0$, and $y_{t} \neq y.$ \\ \hline
\multicolumn{1}{c|}{Optimized Attack} & \multicolumn{1}{l|}{$|\mathcal{T}|>1$.} & \multicolumn{2}{c|}{Non-optimized Attack} & $|\mathcal{T}|=1$. \\ \hline
\multicolumn{1}{c|}{Semantic Attack}                 & \multicolumn{1}{l|}{$\t$ is a semantic part of samples.} & \multicolumn{2}{c|}{Non-semantic Attack} & $\t$ is not a semantic part of samples. \\ \hline
\multicolumn{1}{c|}{Sample-agnostic Attack}                 & \multicolumn{1}{l|}{All $\x'$ contain the same $\bm{t}$.} & \multicolumn{2}{c|}{Sample-specific Attack} & Trigger patterns are sample-specific. \\ \hline
\multicolumn{1}{c|}{Digital Attack} & \multicolumn{1}{l|}{$\x'$ is generated in digital space.} & \multicolumn{2}{c|}{Physical Attack} & Physical space is involved in generating $\x'$. \\ \hline
\multicolumn{1}{c|}{All-to-one Attack}                 & \multicolumn{1}{l|}{All $\x'$ have the same label.} & \multicolumn{2}{c|}{All-to-all Attack} & Different $\x'$ have different labels. \\ \hline
\multicolumn{1}{c|}{White-box Attack} & \multicolumn{1}{l|}{ $\mathcal{D}_t$ is known.} & \multicolumn{2}{c|}{Black-box Attack} & $\mathcal{D}_t$ is unknown. \\ 
\bottomrule
\end{tabular}
\end{table*}

\begin{defn} [Standard, Backdoor, and Perceivable Risk]
\label{def: three risks}
\end{defn}
\begin{itemize}
    \item \emph{The standard risk $R_{s}$ measures whether the infected model $C$ can correctly predict benign samples, $i.e.$, 
    \begin{equation}
    R_{s}(\mathcal{D}) = \mathbb{E}_{(\bm{x}, y) \sim  \mathcal{P}_{\mathcal{D}}} \mathbb{I}\{C(\bm{x}) \neq y\},
    \end{equation}
    where $\mathcal{P}_{\mathcal{D}}$ indicates the distribution behind $\mathcal{D}$ and $\mathbb{I}(\cdot)$ is the indicator function. $\mathbb{I}\{A\}=1$ if and only if the event `$A$' is true.}
    \vspace{0.25em}
    \item \emph{The backdoor risk $R_{b}$ indicates whether backdoor attackers can successfully achieve their malicious purposes in predicting attacked samples, $i.e.$,
    \begin{equation}
        R_{b}(\mathcal{D}) = \mathbb{E}_{(\bm{x}, y) \sim  \mathcal{P}_{\mathcal{D}}} \mathbb{I}\{C(\x') \neq S(y)\},
    \end{equation}
    where $\x' = G_{\t}(\x)$ is the attacked image. 
    }
    \vspace{0.25em}
    \item \emph{The perceivable risk $R_{p}$ denotes whether the poisoned sample is detectable (by human or machine), $i.e.$,
    \begin{equation}
        R_{p}(\mathcal{D}) = \mathbb{E}_{(\bm{x}, y) \sim  \mathcal{P}_{\mathcal{D}}} D(\x'),
    \end{equation}
    where $D(\cdot)$ is an indicator function. $D(\x') = 1$ if and only if $\x'$ can be detected as the malicious sample. 
    }
\end{itemize}

\vspace{1em}
Given a benign training set $\mathcal{D}_t$, existing poisoning-based backdoor attacks can be summarized in a unified framework based on aforementioned definitions, as follows:

\begin{equation} \label{general prob}
    \min_{\t, \boldsymbol{w}} R_{s}(\mathcal{D}_t - \mathcal{D}_{s}) + \lambda_1 \cdot R_{b}(\mathcal{D}_{s}) + \lambda_2 \cdot R_{p}(\mathcal{D}_{s}),
\end{equation}
where $\t \in \mathcal{T}$, $\lambda_1$ and $\lambda_2$ are two non-negative trade-off hyper-parameters, and $\mathcal{D}_{s}$ is a subset of $\mathcal{D}_{t}$. In particular, $\frac{|\mathcal{D}_{s}|}{|\mathcal{D}_{t}|}$ is called \emph{poisoning rate} in existing works.

\vspace{0.4em}
\noindent {\bf Remark}.
Since the indicator function $\mathbb{I}(\cdot)$ used in $R_{s}$ and $R_{b}$ is non-differentiable, it is usually replaced by its surrogate loss (\eg, cross-entropy, KL-divergence) in practice. Besides, as we mentioned, optimization (\ref{general prob}) can reduce to existing attacks through different specifications. For example, when $\lambda_1 = \frac{|\mathcal{D}_{s}|}{|\mathcal{D}_t - \mathcal{D}_{s}|}$, $\lambda_2=0$, and $\t$ is non-optimized (\ie, $|\mathcal{T}|=1$), it reduces to the BadNets \cite{gu2019badnets} and the blended attack \cite{chen2017targeted}; when $\lambda_2=+\infty$ and $D(\x')=\mathbb{I}\{||\x'-\x||_p\leq \epsilon\}$, it reduces to $\epsilon$-bounded invisible backdoor attacks \cite{li2020invisible}. Besides, parameters $\t$ and $\boldsymbol{w}$ could be optimized simultaneously or separately.

\vspace{0.4em}
In particular, this framework can be easily generalized towards other tasks, such as speech recognition, as well. Since there were many different types of tasks and their papers were limited, this generalization is out of the scope of this survey.

\subsection{Evaluation Metrics}
To evaluate the performance of backdoor attacks in the image classification, two classical metrics are usually adopted, including \textbf{(1)} benign accuracy (BA) and \textbf{(2)} attack success rate (ASR), as defined in Section \ref{Sec:def}. The higher the BA and ASR, the better the attack. Besides, the smaller the poisoning rate and the perturbation between the benign image and the poisoned image, the more \emph{stealthy} the attack.

\subsection{Attacks for Image and Video Classification}
\label{sec: attacks for image and video recognition}

\subsubsection{BadNets}
Gu et al. \cite{gu2019badnets} introduced the first backdoor attack in deep learning by poisoning some training samples. This method was called BadNets. Specifically, as demonstrated in Fig. \ref{fig:badnets}, its training process consists of two main parts, including \textbf{(1)} generate some poisoned images via stamping the backdoor trigger onto selected benign images to achieve poisoned sample $(\x', y_t)$, associated with the attacker-specified target label $y_t$, and \textbf{(2)} release the poisoned training set containing both poisoned and benign samples to victims for training their own models. Accordingly, the trained DNN will be infected, which performs well on benign testing samples, similarly to the model trained using only benign samples; however, if the same trigger is contained in an attacked image, then its prediction will be changed to the target label. This attack could happen in all scenarios described in Section \ref{sec:scen} and therefore is a serious security threat. BadNets is the representative of \emph{visible attacks}, which opened the era of this field. Almost all follow-up poisoning-based attacks were carried out based on this method.

\vspace{0.4em}
\subsubsection{Invisible Backdoor Attacks}
Chen et al. \cite{chen2017targeted} first discussed the \emph{invisibility} requirement of poisoning-based backdoor attacks. They suggested that the poisoned image should be indistinguishable compared with its benign version to evade human inspection. To fulfill this requirement, they proposed a \emph{blended strategy}, which generated poisoned images by blending the backdoor trigger with benign images instead of by stamping (as adopted in BadNets \cite{gu2019badnets}). Besides, they showed that even adopting a random noise with a small magnitude as the backdoor trigger can still create the backdoor successfully, which further reduces the risk of being detected.

After that, there was a series of works dedicated to the research of invisible backdoor attacks. In \cite{turner2019label}, Turner et al. proposed to perturb the pixel values of benign images by a backdoor trigger amplitude instead of by replacing the corresponding pixels with the chosen pattern. Zhong et al. \cite{zhong2020backdoor} adopted the universal adversarial attack \cite{moosavi2017universal} to generate backdoor triggers, which minimizes the $\ell^2$ norm of the perturbation to ensure invisibility. After that, \cite{li2020invisible,doan2021lira,doan2021backdoor} proposed to regularize the $\ell^p$ norm of the perturbation when optimizing the backdoor trigger. Liu et al. \cite{liu2020reflection} proposed to adopt a common phenomenon ($i.e.$, the reflection) as the trigger for stealthiness. Nguyen et al. \cite{nguyen2020wanet} adopted warping-based triggers, which are more invisible for human inspection. Recently, \cite{bagdasaryan2020blind} viewed the backdoor attack as a special multi-task learning, where they fulfilled the invisibility through poisoning the loss computation. Cheng et al. \cite{cheng2021deep} proposed to conduct the invisible attack in the feature space via style transfer. Different from previous works whose poisoned samples were generated in the pixel domain, \cite{hammoud2021check,wang2021backdoor} generated invisible trigger patterns in the frequency domain. Most recently, Li et al. \cite{li2021invisible} adopted DNN-based image steganography to generate invisible backdoor triggers. Compared with previous methods, this attack is not only invisible but can also bypass most existing backdoor defenses, since its trigger patterns are sample-specific.

\begin{figure*}[ht]
	\centering
	\includegraphics[width=0.98\textwidth]{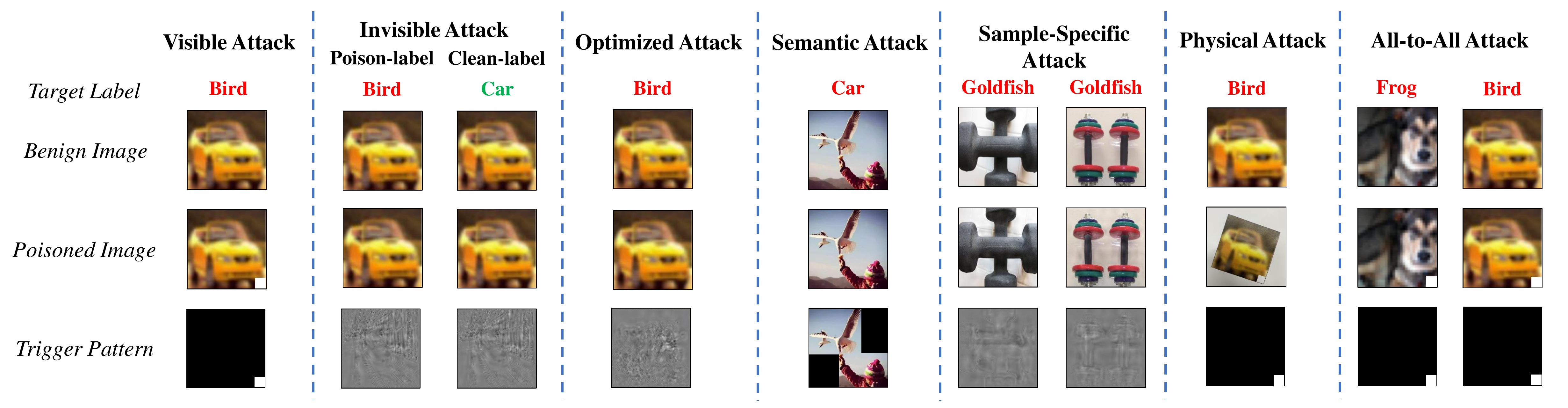}
	\vspace{-0.4em}
	\caption{An example of poisoned samples generated by different types of backdoor attacks. \textbf{(1)} In the visible attack, the backdoor trigger is a white-square stamped on the bottom right corner of the poisoned image, which is visible. \textbf{(2)} In the invisible attack, the trigger is a noise with a small magnitude, which is invisible. Moreover, the target label of the poisoned image is different from the ground-truth label of its benign version in the poison-label attack, whereas these labels are the same in the clean-label attack. \textbf{(3)} In the optimized attack, the trigger is optimized through the targeted universal adversarial attack associated with the target class instead of a simple handcraft pattern. \textbf{(4)} The poisoned image is exactly the same as its benign version in the semantic attack. In this case, the trigger is the combination of two semantic objects ($i.e.$, `bird' and `human'). Images containing these objects simultaneously will be classified by the infected models as the `car'. \textbf{(5)} In the sample-specific attack, the trigger patterns are sample-specific instead of sample-agnostic. \textbf{(6)} In the physical attack, the (digital) poisoned image is captured by the camera from the physical space. \textbf{(7)} Different from all-to-one attacks where all poisoned samples have the same target label, different poisoned samples may have different target labels in the all-to-all attack.
	} 
	\label{fig:poisonedsamples}
	\vspace{-0.8em}
\end{figure*}

Although a poisoned image is similar to its benign version in invisible attacks, however, its source label is usually different from the target label. In other words, all those methods are \emph{poison-label invisible attacks}, where the poisoned samples seem to be mislabeled. Accordingly, an invisible attack still could be detected by examining the image-label relationship of training samples. To address this problem, a special sub-class of invisible poisoning-based attacks, dubbed \emph{clean-label invisible attacks}, was proposed. It is more serious and therefore worth more attention. Turner et al. \cite{turner2019label} first explored the clean-label attack, where they leveraged adversarial perturbations or generative models to first modify some benign images from the target class and then conducted the standard invisible attack. The modification process is to alleviate the effects of `robust features' contained in the poisoned samples to ensure that the trigger can be successfully learned by the DNNs. Recently, Zhao et al. \cite{zhao2020clean} extended this idea in attacking video classification, where they adopted universal perturbation instead of a given one as the trigger pattern. Quiring et al. \cite{quiring2020backdooring} proposed to conceal the trigger via image-scaling attacks \cite{xiao2019seeing}. Another interesting clean-label attack method is to inject the information of a poisoned sample generated by a previous visible attack into the texture of an image from the target class by minimizing their distance in the feature space, as suggested in \cite{saha2020hidden}. Following the settings in \cite{saha2020hidden}, Souri et al. \cite{souri2021sleeper} formulated the backdoor attacks as a bi-level optimization \cite{liu2021investigating}, based on which they proposed a new clean-label backdoor attack. Most recently, Shumailov et al. \cite{shumailov2021manipulating} proposed to inject hidden backdoors via manipulating the order of training samples without changing samples.

In particular, clean-label backdoor attacks usually suffered from low attack effectiveness compared with poison-label attacks, although they are more stealthy. How to balance the stealthiness and effectiveness of attacks is still an open question and worth further explorations.

\vspace{0.4em}
\subsubsection{Optimized Backdoor Attacks}
Triggers are the core of poisoning-based attacks. As such, analyzing how to design a better trigger instead of simply using a given non-optimized patch is of great significance and has attracted some attention. In general, backdoor attacks can be formulated as a bi-level optimization \cite{liu2021investigating}, $i.e.$, $\min_{\boldsymbol{w}} R_{s}(\mathcal{D}_t - \mathcal{D}_{s};\boldsymbol{w}) + \lambda_1 \cdot R_{b}(\mathcal{D}_{s};\t^{*},\boldsymbol{w}) + \lambda_2 \cdot R_{p}(\mathcal{D}_{s};\t^{*},\boldsymbol{w})$, $s.t.$, $\t^{*} = \min_t R_{b}(\mathcal{D}_{s};\t,\boldsymbol{w})$. Optimized attacks generated poisoned samples with optimized triggers to achieve better performance. To the best of our knowledge, Liu et al. \cite{liutrojaning} first explored this problem, where they proposed to optimize the trigger so that the important neurons can achieve the maximum values. After that, with the hypothesis that if a perturbation can induce most samples toward the decision boundary of the target class then it will serve as an effective trigger, \cite{zhong2020backdoor,zhao2020clean,garg2020can} proposed to generate trigger through universal adversarial perturbation. These methods can be regarded as the heuristic solutions of the aforementioned bi-level optimization. Recently, \cite{li2020invisible,doan2021lira,doan2021backdoor,souri2021sleeper} solved the bi-level optimization problem directly. For example, \cite{li2020invisible,doan2021lira,doan2021backdoor} alternately optimized the upper-level and lower-level sub-problems while \cite{souri2021sleeper} adopted the gradient matching \cite{geiping2020witches}. However, optimized backdoor attacks usually suffer from poor generalization, $i.e.$, overfits to a particular model structure or model status. Although existing works introduced model-ensemble or carefully designed the alternately optimization process to alleviate the overfitting, how to better balance the effectiveness and generalization of the optimized triggers is still an important open question.

\vspace{0.4em}
\subsubsection{Semantic Backdoor Attacks}
The majority of backdoor attacks, $i.e.$, the \emph{non-semantic attacks}, assume that backdoor triggers are independent of benign images. As such, attackers need to modify the image in the digital space to activate hidden backdoors in the inference process. Is it possible that a semantic part of samples can also serve as the trigger pattern, such that the attacker is not required to modify the input at inference time to deceive the infected model? 
Bagdasaryan et al. first explored this problem and proposed a novel type of backdoor attacks \cite{bagdasaryan2020backdoor,bagdasaryan2020blind}, $i.e.,$ the \emph{semantic backdoor attacks}. Specifically, they demonstrated that assigning an attacker-chosen label to all images with certain features, \eg, green cars or cars with racing stripes, for training can create semantic backdoors in the infected DNNs. Accordingly, the infected model will automatically misclassify testing images containing pre-defined semantic information without any image modification. A similar idea was also explored in \cite{lin2020composite}, where the hidden backdoor can be activated by the combination of certain objects in the image. Since these attacks do not require modifying images in the digital space, they are more malicious and worth further explorations.

\vspace{0.4em}
\subsubsection{Sample-specific Backdoor Attacks}
Currently, almost all backdoor attacks were sample-agnostic, $i.e.$, all poisoned samples contained the same trigger pattern. This property was widely used in the design of backdoor defenses, such as trigger synthesis based defenses \cite{wang2019neural,chen2019deepinspect,guo2019tabor,shen2021backdoor,dong2021black} and saliency-based defenses \cite{huang2019neuroninspect,chou2020sentinet}. Nguyen et al. \cite{nguyen2020input} proposed the first sample-specific backdoor attack, where different poisoned samples contain different trigger patterns. This attack bypassed many existing backdoor defenses for it broke their fundamental assumptions. However, it needs to control the training loss except for solely modifying training samples and their triggers are still visible, which significantly reduces its threat in real-world applications. After that, Li et al. \cite{li2021invisible} proposed the first poison-only sample-specific backdoor attack with invisible trigger patterns, inspired by the advanced DNN-based image steganography. A similar idea is also explored in \cite{zhang2021poison}, where they embedded trigger patterns in the edge structure of poisoned images. Since these attacks can bypass most existing backdoor defenses, they pose a serious security threat and therefore worth further explorations.

\vspace{0.4em}
\subsubsection{Physical Backdoor Attacks}
Different from previous \emph{digital attacks} where attacks were conducted completely in the digital space, the physical space was also involved when generating poisoned samples in the \emph{physical attacks}. Chen et al. \cite{chen2017targeted} first explored the landscape of this attack, where they adopted a pair of glasses as the physical trigger to mislead the infected face recognition system developed in a camera. Further exploration of attacking face recognition in the physical world was also discussed by Wenger et al. \cite{wenger2020backdoor}. A similar idea was also discussed in \cite{gu2019badnets}, where a post-it note was adopted as the trigger in attacking traffic sign recognition deployed in the camera. Recently, Li et al. \cite{li2021backdoor} demonstrated that existing digital attacks fail in the physical world since the involved transformations (\eg, rotation, and shrinkage) change the location and appearance of triggers in attacked samples. This inconsistency will greatly reduce the performance of backdoor attacks. Based on this understanding, they proposed a transformation-based attack enhancement so that the enhanced attacks remain effective in the physical world. This attempt is an important step towards successful backdoor attacks in real-world applications.

\vspace{0.4em}
\subsubsection{All-to-all Backdoor Attacks}
Based on the type of target labels, existing backdoor attacks can be divided into two main categories, including the \emph{all-to-one attacks} and the \emph{all-to-all attacks}. Specifically, all-to-one attacks assumed that all poisoned samples have the same target label no matter what their ground-truth label is, $i.e.$, $S(y) = y_t, \forall y \in \{1, \cdots, K\}$. In contrast, different poisoned samples may have different labels in all-to-all attacks. For example, the label shifting function was assigned as $S(y)= (y+1) \mod K$ in \cite{gu2019badnets,nguyen2020input,doan2021lira}. The all-to-all attacks can bypass many target-oriented defenses ($e.g.$, \cite{wang2019neural,chou2020sentinet,dong2021black}) for its complicated target shifting and therefore is more serious compared with the all-to-one attacks. However, there were only a few studies in all-to-all attacks. How to better design the all-to-all attack and the analysis of its properties remain blank.

\vspace{0.4em}
\subsubsection{Black-box Backdoor Attacks}
Different from previous \emph{white-box attacks} which required to access the training samples, \emph{black-box attacks} adopted the settings that the training set is inaccessible. In practice, the training dataset is usually not shared due to privacy or copyright concerns, therefore black-box attacks are more realistic than white-box ones. In general, black-box backdoor attackers generated some substitute training samples at first. For example, in \cite{liutrojaning}, attackers generated some representative images of each class by optimizing images initialized from another dataset such that the prediction confidence of the selected class reaches maximum. With the substitute training samples, white-box attacks can be adopted for backdoor injection. Black-box backdoor attacks are significantly difficult than white-box ones and there were only a few works in this area.


\subsection{Attacks against Other Fields or Paradigms}

Currently, most existing backdoor attacks against other tasks or paradigms were still poisoning-based. Accordingly, except for task-specific requirements, most methods focused on \textbf{(1)} how to design the trigger, \textbf{(2)} how to define the attack stealthiness, and \textbf{(3)} how to bypass potential defenses. The huge differences between different tasks and paradigms make the answers to the above questions completely different. For example, the stealthiness in image-related tasks can be defined as the pixel-wise distance ($e.g.$, $\ell^p$ norm) between the poisoned sample and its benign version; however, in natural language processing (NLP), changing even a word or character may still make the modification visible to human since it may cause grammar or spelling errors. 

Natural language processing is currently the most extensive research field in backdoor attacks besides image classification. In \cite{dai2019backdoor}, Dai et al. discussed how to attack against LSTM-based sentiment analysis. Specifically, they proposed a BadNets-like approach, where an emotionally neutral sentence was used as the trigger and was randomly inserted into some benign training samples. In \cite{chen2020badnl}, Chen et al. further explored this problem, where three different types of triggers (\ie, char-level, word-level, and sentence-level triggers) were proposed and reached decent performance. Besides, Kurita et al. \cite{kurita2020weight} demonstrated that sentiment classification, toxicity detection, and spam detection can also be attacked even after fine-tuning. Most recently, other backdoor attacks were also introduced, targeting different trigger types \cite{qi2021turn,qi2021hidden,yang2021rethinking,qi2021mind} and model components \cite{yang2021careful,li2021backdoora} in different NLP tasks. Except for NLP-related tasks, researchers also revealed the backdoor threats in graph neural networks (GNN) \cite{zhang2020backdoor,xi2020graph,chen2021dyn}, 3D point cloud \cite{tian2021poisoning,xiang2021backdoor,li2021pointba}, semi-/self-supervised learning \cite{yan2021deep,jia2022badencoder,carlini2022poisoning}, reinforcement learning \cite{kiourti2020trojdrl,wang2021backdoorl,ashcraft2021poisoning}, model quantization \cite{ma2021quantization,hong2021qu,pan2021understanding}, acoustics signal processing \cite{zhai2021backdoor,koffas2021can}, malware detection \cite{li2021backdoormal,severi2021explanation}, and others \cite{li2021hidden,fang2022backdoor,li2022few}.


Except for the classical training paradigm, how to backdoor collaborative learning, especially federated learning, have attracted the most attention. In \cite{bagdasaryan2020backdoor}, Bagdasaryan et al. introduced the first backdoor attack against federated learning by amplifying the poisoned gradient of node servers. After that, Bhagoji et al. \cite{bhagoji2019analyzing} discussed the stealthy model-poisoning backdoor attack, and Xie et al. \cite{xie2019dba} introduced a distributed backdoor attacks against the federated learning. 
Most recently, \cite{wang2020attack} theoretically verified that backdoor attacks are unavoidable if a model is vulnerable to adversarial examples under mild conditions in federated learning. Besides, the backdoor attacks towards meta federated learning \cite{chen2020backdoor} and feature-partitioned collaborative learning \cite{liu2020backdoor,liu2022defending} were also discussed. In contrast, some works \cite{sun2019can,xie2021crfl,safa2021defending,liu2021privacy} also questioned whether federal learning is really easy to be attacked. 
Except for collaborative learning, the backdoor threat towards another important learning paradigm, $i.e.$, the transfer learning, was also discussed in \cite{yao2019latent,wang2020backdoor,chen2022badpre}.

\begin{table*}[ht]
\centering
\vspace{-0.5em}
\caption{Comparison among the backdoor attack, adversarial attack, and data poisoning. }
\vspace{-0.5em}
\scalebox{0.93}{
\begin{tabular}{lllll}
\toprule
\textbf{Attack Category} & \textbf{Attacker's Goals} & \textbf{Attack Mechanism} & \textbf{Training Capacities} & \textbf{Inference Capacities} \\ \midrule
Backdoor Attack    & \makecell[l]{Misclassify (modified) attacked samples; \\ Behave normally on benign samples.} & \makecell[l]{Excessive learning ability \\ of models. } & Under control. & Out of control. \\ \hline
Adversarial Attack & \makecell[l]{Misclassify (modified) attacked samples; \\ Behave normally on benign samples.} & \makecell[l]{Behavior differences \\ between models and humans.} & Out of control. & \makecell[l]{Query the model \\ multiple times \\to generate adversarial\\ perturbations by optimization.} \\ \hline 
Classical Data Poisoning & Reduce model generalization. & \makecell[l]{The sensitiveness of\\ training process.} & \makecell[l]{Can only modify \\ the training set.} & Out of control. \\  \hline
Advanced Data Poisoning & \makecell[l]{Misclassify (unmodified) targeted samples; \\ Behave normally on benign samples. } & \makecell[l]{Excessive learning ability \\ of models. } & \makecell[l]{Can only modify \\ the training set.} & Out of control.
\\ \bottomrule
\end{tabular}
}
\vspace{-0.5em}
\label{tab:compar}
\end{table*}

\subsection{Backdoor Attacks for Positive Purposes}
Except for malicious applications, how to use backdoor attacks for positive purposes has also obtained some preliminary explorations. Adi et al. \cite{adi2018turning} adopted backdoor attacks in defending against model stealing via ownership verification. Specifically, they proposed to watermark the DNNs through backdoor embedding, which can be used to examine the model ownership. However, a recent study \cite{li2022defending} revealed that this approach could fail, especially when it is complicated, since the stealing process may change or even remove hidden backdoors contained in the victim models. Besides, Sommer et al. \cite{sommer2020towards} discussed how to verify whether the server truly erases their data when users require data deletion through poisoning-based backdoor attacks. Specifically, under their settings, each user can poison part of its data with a specific trigger and target label. Accordingly, each user can leave a unique trace in the server for deletion verification after the server is trained on user data. Besides, Shan et al. \cite{shan2020using} introduced a trapdoor-enabled adversarial defense, where the hidden backdoor was injected by the defender to prevent attackers from discovering the real weakness in a model. The motivation was that the generated adversarial perturbation towards an infected model will converge near the trapdoor pattern, which was easily detected by the defender. Moreover, Li et al. \cite{li2020open} discussed how to protect open-sourced datasets based on backdoor attacks. Specifically, they formulated this problem as determining whether the dataset has been adopted to train a suspicious model. They proposed a hypothesis test based method for the verification, based on the posterior probability of the benign samples and their attacked version generated by the suspicious model. Most recently, backdoor attacks were also adopted for interpreting DNNs \cite{zhao2021deep} and the evaluation of explainable AI methods \cite{lin2020you}.

\vspace{0.8em}
\section{Non-poisoning-based Backdoor Attacks}
\label{sec:nonpoisoning-based-attack}
Except for poisoning-based backdoor attacks, recent literature also proposed some non-poisoning-based attacks. These methods embedded hidden backdoors not directly based on data poisoning during the training process. For example, attackers may directly change model weights or even the model structure without the training process. Their existence demonstrates that backdoor attacks could also happen at other stages ($e.g.,$ deployment stage) instead of simply the data collection or training stages, which further reveals the severity of backdoor threats.

\subsection{Weights-oriented Backdoor Attacks}
In the weights-oriented backdoor attacks, attackers modified model parameters directly instead of through training with poisoned samples. To the best of our knowledge, Dumford et al. \cite{dumford2018backdooring} proposed the first weights-oriented attack where they adopted a greedy search across models with different perturbations applied to a pre-trained model’s weights. It is also the first non-poisoning-based backdoor attack. After that, Rakin et al. \cite{rakin2020tbt} introduced a bit-level weights-oriented backdoor attack, $i.e.$, the targeted bit trojan (TBT), which flipped critical bits of weights stored in the memory. The proposed method achieved remarkable performance, where attackers were able to mislead ResNet-18 \cite{he2016deep} on the CIFAR-10 dataset \cite{CIFAR} with 84 bit-flips out of 88 million weight bits. A similar idea was also introduced in \cite{chen2021proflip}, where attackers can significantly reduce the required flipping bits to embed hidden backdoors. Besides, Garg et al. \cite{garg2020can} proposed to add adversarial perturbations on the model parameters of the benign model for injecting backdoors, showing a novel security threat of using publicly available trained models. Most recently, Zhang et al. \cite{zhang2021inject} formulated the behavior of maintaining accuracy on benign samples as the consistency of infected models and provided a theoretical explanation of the adversarial weight perturbation (AWP) in backdoor attacks. Based on the analysis, they also introduced a new AWP-based backdoor attack with better global and instance-wise consistency.

Different from previous approaches where the backdoor is embedded in the parameters directly, Guo et al. \cite{guo2020trojannet} proposed TrojanNet to encode the backdoor in the infected DNNs activated through a secret weight permutation. Specifically, training a TrojanNet is similar to the \emph{multi-task learning}, although the benign task and malicious task share no common features. Besides, the authors also proved that the decision problem to determine whether the model contains a permutation that triggers the hidden backdoor is NP-complete, and therefore the backdoor detection is almost impossible.

\subsection{Structure-modified Backdoor Attacks}
Structure-modified backdoor attacks injected hidden backdoors into benign models by changing their model structures. These attacks could happen when using third-party models or in the deployment stage. To the best of our knowledge, Tang et al. \cite{tang2020embarrassingly} proposed the first structure-modified attack, where they inserted a trained malicious backdoor module ($i.e.$, a sub-DNN) into the target model for embedding hidden backdoors. This attack was simple yet effective and the malicious module can be combined with all DNNs. A similar idea was also explored in \cite{li2021deeppayload}, where attackers embedded malicious conditional logics into the target DNNs by adding malicious payload containing the conditional module and the trigger detector. Most recently, Qi et al. \cite{qi2021subnet} proposed to directly replace instead of adding a narrow subnet of the benign model to conduct the backdoor attack. This method is effective in both digital and physical scenarios.

\comment{
\begin{figure*}[ht]
	\centering
	\includegraphics[width=0.97\textwidth]{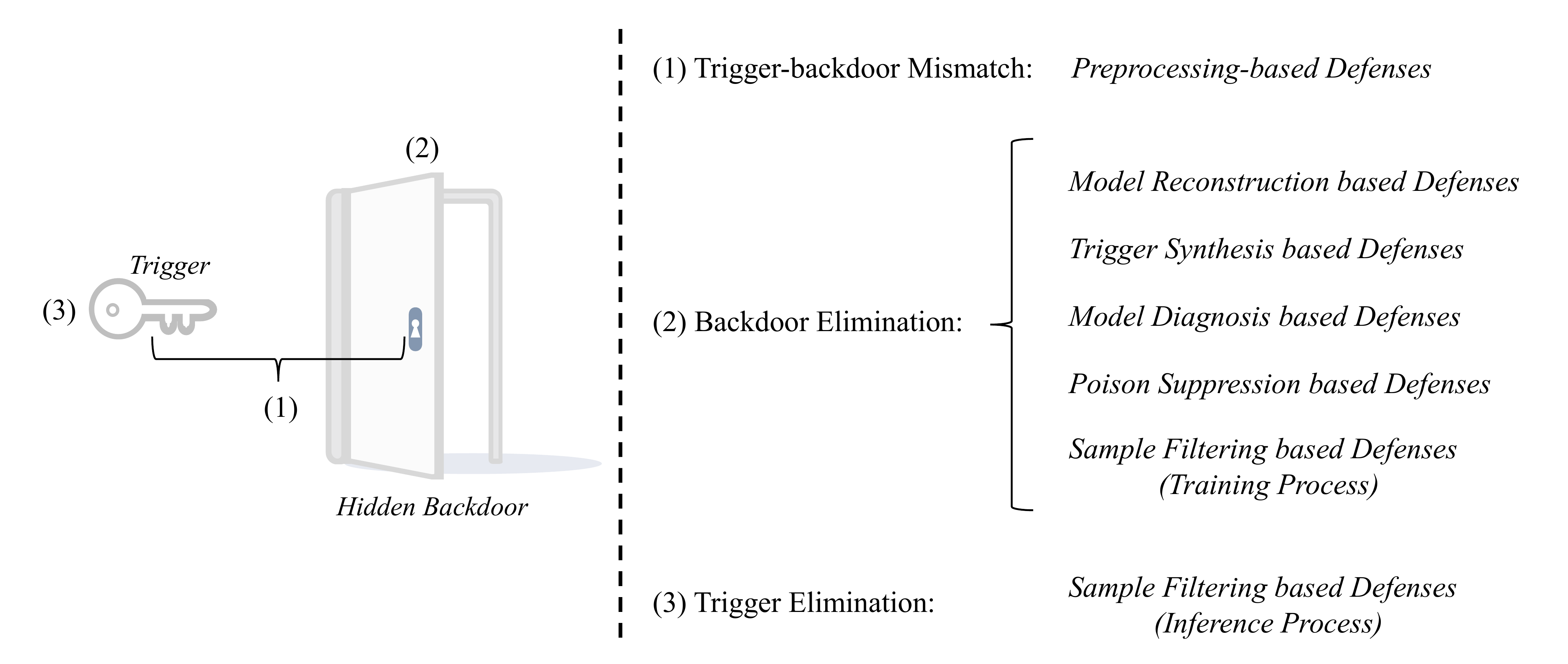}
	\caption{An illustration of empirical backdoor defenses. Intuitively, poisoning-based backdoor attack is similar to unlock a door with the corresponding key. Accordingly, three main paradigms, including \textbf{(1)} trigger-backdoor mismatch, \textbf{(2)} backdoor elimination, and \textbf{(3)} trigger elimination, can be adopted to defend the attack. Different types of approaches were proposed towards aforementioned paradigms.}
	\label{fig:defenses}
\end{figure*}
}

\section{Connection with Related Realms}\label{sec:discussion}
In this section, we discuss the similarities and differences between backdoor attacks and related realms. Those connections are summarized in Table \ref{tab:compar}.

\subsection{Backdoor Attacks and Adversarial Attacks}
Both adversarial attacks and backdoor attacks modify the benign testing samples to make models misbehave during the inference process. Especially when the adversarial perturbations are sample-agnostic in universal adversarial attacks \cite{moosavi2017universal, mopuri2018generalizable, thys2019fooling}, these attacks seem to be the same. As such, researchers who are not familiar with the backdoor attack may question its research significance for it requires additional controls of the training process to some extent.

\begin{figtab*}
  \begin{minipage}[tb]{0.32\linewidth}
    \centering
    \includegraphics[width = \linewidth]{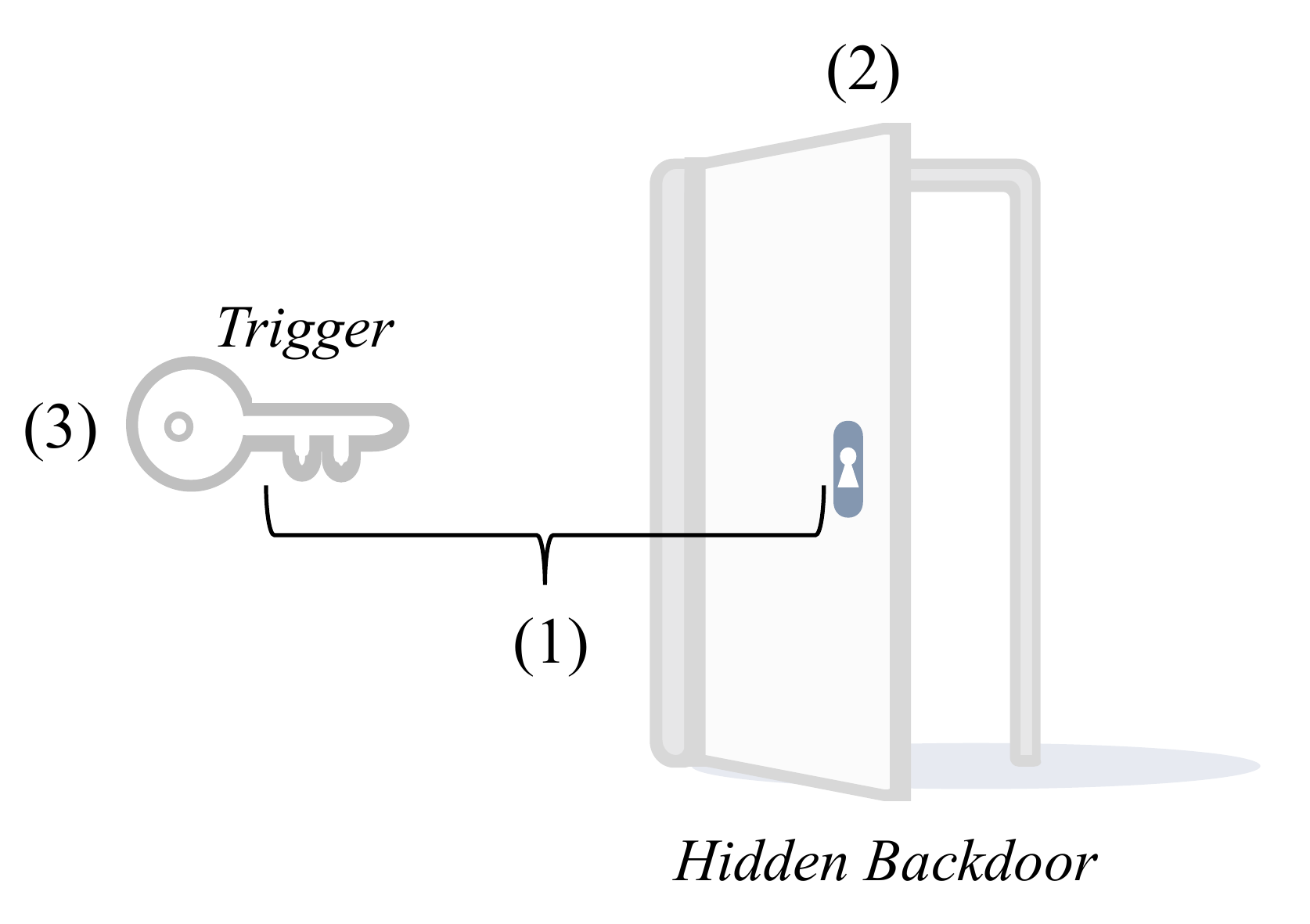}
    \vspace{-2em}
    \figcaption{An illustration of backdoor attacks and three corresponding defense paradigms. Intuitively, the poisoning-based backdoor attack is similar to unlock a door with the corresponding key. Accordingly, three main paradigms, including \textbf{(1)} trigger-backdoor mismatch, \textbf{(2)} backdoor elimination, and \textbf{(3)} trigger elimination, can be adopted to defend the attack. Different types of approaches were proposed towards the aforementioned paradigms, as illustrated in Table \ref{tab:defense}.}
	\label{fig:defenses}
  \end{minipage}\quad
  \begin{minipage}[tb]{0.66\linewidth}
    \centering
    \footnotesize
    \tabcaption{Summary of existing empirical backdoor defenses. In particular, since some literature proposed different types of defenses simultaneously, they will appear multiple times in this table.}
\scalebox{0.92}{    
\begin{tabular}{c|l|l}
\toprule
\textbf{Defense Paradigm} & \textbf{Defense Sub-category} & \textbf{Literarure} \\ \hline
Trigger-backdoor Mismatch 
& Preprocessing-based Defenses & \tabincell{l}{\cite{liu2017neural, doan2019februus, udeshi2019model,villarreal2020confoc}\\ \cite{li2021backdoor,zeng2020deepsweep}}  \\ \hline
\multirow{9}{*}{Backdoor Elimination} 
& Model Reconstruction based Defenses & \tabincell{l}{\cite{liu2017neural, liu2018fine, zhao2020bridging,yoshida2020disabling}\\ \cite{li2021neural,wu2021adversarial,zeng2022adversarial}} \\ \cline{2-3}
& Trigger Synthesis based Defenses & \tabincell{l}{\cite{wang2019neural,chen2019deepinspect,qiao2019defending,guo2019tabor}\\ \cite{zhu2020gangsweep,cheng2019defending,aiken2020neural,harikumar2020scalable}\\ \cite{xiang2020detection,shen2021backdoor,dong2021black,guo2021aeva}\\
\cite{hu2022trigger}}  \\ \cline{2-3} & Model Diagnosis based Defenses &\tabincell{l}{ \cite{huang2019neuroninspect, xu2019detecting,  kolouri2020universal,huangone20} \\
\cite{wang2020practical,zheng2021topological,xiang2022post}}  \\ \cline{2-3}  
& Poison Suppression based Defenses &  \tabincell{l}{\cite{du2020,hong2020effectiveness,borgnia2021strong,liu2021removing}\\
\cite{li2021anti,huang2022backdoor}} \\ \cline{2-3}  
& Training Sample Filtering based Defenses & \tabincell{l}{\cite{tran2018spectral,chen2019detecting,tang2019demon,soremekun2020exposing}\\ \cite{chan2019poison,chou2020sentinet,hayase2021spectre,wang2021unified}\\
\cite{zeng2021rethinking}} \\ \hline
Trigger Elimination  
& Testing Sample Filtering based Defenses & \tabincell{l}{\cite{gao2019strip, subedar2019deep,du2020,jin2020unified}\\ \cite{javaheripi2020cleann}} \\ \bottomrule
\end{tabular}
}
\label{tab:defense}
  \end{minipage}
\end{figtab*}

However, these attacks still have essential differences although they enjoy certain similarities. {\bf (1)} From the aspect of the attacker's capacity, adversarial attackers need to control the inference process (to a certain extent) but not the training process of models. Specifically, they need to query the model results or even gradients multiple times to generate adversarial perturbations by optimization given a fixed targeted model. In contrast, backdoor attackers require to modify some training stages ($e.g.$, data collection, model training) without any additional requirements in the inference process. {\bf (2)} From the perspective of attacked samples, the perturbation is known ($i.e.$, non-optimized) by backdoor attackers whereas adversarial attackers need to obtain it through the optimization process based on the output of the model. Such optimization in adversarial attacks requires multiple queries \cite{dong2019efficient,chen2020boosting,tramer2020adaptive}. As such, adversarial attacks are unable to be real-time in many cases for the optimization process takes time. {\bf (3)} Their mechanism also has essential differences. Adversarial vulnerability results from the differences in behaviors of models and humans. In contrast, backdoor attackers utilize the excessive learning ability of DNNs to build a latent connection between the trigger patterns and the target labels.

Most recently, there were also a few works studying the latent connection between adversarial attacks and backdoor attacks. For example, Weng et al. \cite{weng2020trade} empirically demonstrated that defending against adversarial attacks via adversarial training may increase the risks of backdoor attacks.

\vspace{-0.3em}
\subsection{Backdoor Attacks and Data Poisoning}
In general, there are two types of data poisoning, including the classical and the advanced one. The former one intends to reduce model generalization, $i.e.,$ letting the infected models behave well on training samples whereas having bad performance on testing samples. In contrast, advanced data poisoning makes infected models behave well on testing samples whereas having bad performance on some attacker-specified targeted samples which are not contained in the training set.

Data poisoning and (poisoning-based) backdoor attacks share many similarities in the training phase. In general, they all aim at misleading models in the inference process by introducing poisoned samples during the training process. However, they also have many intrinsic differences.

Firstly, compared with classical data poisoning, backdoor attacks preserve the performance of predicting benign samples. In other words, backdoor attacks have different attacker's goals compared with classical data poisoning. Besides, these attacks have different mechanisms. Specifically, the effectiveness of classical data poisoning is mostly due to the sensitiveness of the training process so that even a small domain shift of training samples may lead to significantly different decision surfaces of infected models. Moreover, backdoor attacks are also more stealthy than classical data poisoning. Users can easily detect classical data poisoning by evaluating the performance of trained models on a local verification set, while this method has limited benefits in detecting backdoor attacks. Secondly, backdoor attacks are also different from advanced data poisoning. Specifically, there is no trigger in advanced data poisoning, which does not require modifying targeted samples during the inference process. Correspondingly, advanced data poisoning can only misclassify (a few) specific samples, which limits its threats in many scenarios.

In particular, the studies of existing data poisoning have also inspired the research on backdoor learning due to their similarities. For example, Hong et al. \cite{hong2020effectiveness} demonstrated that the defense towards data poisoning may also have benefits in defending backdoor attacks, as illustrated in Section \ref{sec:PoisonSuppression}.

\section{Backdoor Defenses}\label{sec:defense}
To alleviate the backdoor threats, several backdoor defenses were proposed. Existing methods mostly aim at defending against poisoning-based attacks and can be divided into two main categories, including \emph{empirical backdoor defenses} and \emph{certified backdoor defenses}. Specifically, empirical defenses were proposed based on some understandings of existing attacks and had decent performances in practice, whereas their effectiveness has no theoretical guarantee. In contrast, the validity of certified backdoor defenses is theoretically guaranteed under certain assumptions, whereas it is generally weaker than that of empirical defenses in practice. At present, certified defenses are all based on the \emph{random smoothing} \cite{cohen2019certified}, while empirical ones have multiple types of approaches.

\subsection{Empirical Backdoor Defenses}

Intuitively, poisoning-based backdoor attacks are similar to unlock a door with the corresponding key. In other words, there are three indispensable requirements to ensure the success of backdoor attacks, including \textbf{(1)} having a hidden backdoor in the (infected) model, \textbf{(2)} containing triggers in (attacked) samples, and \textbf{(3)} the trigger and the backdoor are matched, as shown in Fig. \ref{fig:defenses}. Accordingly, three main defense paradigms, including \textbf{(1)} trigger-backdoor mismatch, \textbf{(2)} backdoor elimination, and \textbf{(3)} trigger elimination, can be adopted to defend existing attacks. Different types of approaches were proposed towards the aforementioned paradigms, which are summarized in Table \ref{tab:defense} and will be further demonstrated as follows:

\vspace{0.4em}
\subsubsection{Preprocessing-based Defenses}
These methods introduce a preprocessing module before feeding samples into DNNs to change trigger patterns contained in attacked samples. Accordingly, the modified triggers no longer match the hidden backdoor and therefore preventing backdoor activation.

Liu et al. \cite{liu2017neural} proposed the first preprocessing-based backdoor defense, where they adopted a pre-trained auto-encoder as the preprocessor. Inspired by the idea that the trigger regions contributed most to the prediction, Doan et al introduced a two-stage image preprocessing approach ($i.e.$, Februus) in \cite{doan2019februus}. At the first stage, Februus used GradCAM \cite{selvaraju2017grad} to identify influential regions, which will then be removed and replaced by a neutralized-color box. After that, Februus adopted a GAN-based inpainting method to reconstruct the masked regions to alleviate its adverse effects ($e.g.$, benign accuracy drop). After that, Udeshi el al. \cite{udeshi2019model} used the dominant color in the image to make a square-like trigger blocker in the preprocessing stage, which was adopted to locate and remove the backdoor trigger. This approach was motivated by the understanding that placing a trigger blocker at the position of trigger patterns in attacked images will significantly change model predictions. Vasquez et al. \cite{villarreal2020confoc} proposed to preprocess the image through style transfer. Recently, Li et al. \cite{li2021backdoor} discussed the property of existing poisoning-based attacks with static trigger patterns. They demonstrated that if the \emph{appearance} or \emph{location} of the trigger is slightly changed, the attack performance may degrade sharply. Based on this observation, they proposed to adopt spatial transformations ($e.g.$, shrinking, flipping) for the defense. Compared with previous methods, this method is more efficient since it requires almost no additional computational costs. A similar idea was explored in \cite{zeng2020deepsweep}, where they introduced and evaluated more transformations in both fine-tuning and inference processes.

\vspace{0.4em}
\subsubsection{Model Reconstruction based Defenses}
Different from preprocessing-based defenses, model reconstruction based methods aim at removing hidden backdoors in the infected model by modifying suspicious models directly. As such, even if the trigger is contained in attacked samples, the reconstructed model will still correctly predict them since the hidden backdoors were already removed.

Liu et al. \cite{liu2017neural} proposed to retrain the trained suspicious model with some local benign samples to reduce backdoor threats. Its effectiveness is mostly due to the \emph{catastrophic forgetting} of DNNs \cite{kirkpatrick2017overcoming}, $i.e.$, the hidden backdoor is gradually removed as the training goes since the retraining set contains no poisoned samples. 
This idea was further explored by Zeng et al. \cite{zeng2022adversarial}, where they formulated the retraining as a mini-max problem and adopted the implicit hyper-gradients to account for the interdependence between inner and outer optimization. Motivated by the observation that the backdoor-related neurons are usually dormant when predicting benign samples, Liu et al. \cite{liu2018fine} proposed to prune those neurons to remove the hidden backdoor. Specifically, they proposed a ﬁne-pruning method, which ﬁrst prunes the DNNs and then ﬁne-tunes the pruned network to combine the beneﬁts of the pruning and ﬁne-tuning defenses. A similar idea was further explored in \cite{wu2021adversarial}, where they used adversarial weight perturbation to amplify the differences between benign and malicious neurons. In \cite{zhao2020bridging}, Zhao et al. showed that the hidden backdoor of infected DNNs can be repaired based on the \emph{mode connectivity} technique \cite{garipov2018loss} with a certain amount of benign samples. Most recently, Yoshida et al. \cite{yoshida2020disabling} and Li et al. \cite{li2021neural} adopted \emph{knowledge distillation} technique \cite{hinton2015distilling} to reconstruct (infected) DNNs, based on the understanding that the distillation process perturbs backdoor-related neurons and therefore can remove hidden backdoors.

\vspace{0.4em}
\subsubsection{Trigger Synthesis based Defenses}
Except for eliminating hidden backdoors directly, trigger synthesis based defenses first synthesize the backdoor trigger, followed by the second stage that the hidden backdoor is eliminated by suppressing trigger's effects. These defenses enjoy certain similarities with reconstruction-based ones in the second stage. For example, pruning and retraining are the common techniques used in removing the hidden backdoor in both types of defenses. However, compared with the reconstruction-based defenses, the trigger information obtained in synthesis-based defenses makes the removal process more effective and efficient.

To the best of our knowledge, Wang et al. \cite{wang2019neural} proposed the first trigger synthesis based defense ($i.e.$ Neural Cleanse), where defenders first obtained potential trigger patterns towards every class and then determined the final synthetic trigger and its target label based on anomaly detection. A similar idea was also discussed in \cite{chen2019deepinspect,harikumar2020scalable,xiang2020detection,xiang2022post} where they designed different trigger reversion or detection techniques. Qiao et al. \cite{qiao2019defending} noticed that the reversed trigger synthesized by Neural Cleanse is usually signiﬁcantly different from that was used in the training process, inspired by which they first discussed the generalization of the backdoor trigger. They demonstrated that infected models will generalize their original triggers during the training process. Accordingly, they proposed to recover the trigger distribution rather than a specific trigger for the defense, based on a max-entropy staircase approximator. A similar idea was also discussed in \cite{zhu2020gangsweep}, where they proposed a GAN-based method to synthesize trigger distribution. In \cite{guo2019tabor}, they showed that the detection process used for determining the synthetic trigger in \cite{wang2019neural} suffers from several failure modes, based on which they proposed a new defense method. Besides, Cheng et al. \cite{cheng2019defending} revealed that the $\ell^\infty$ norm of the activation values can be used to distinguish backdoor related neurons based on the synthetic trigger. Accordingly, they proposed to perform $\ell^\infty$-based neuron pruning to remove neurons with high activation values in response to the trigger. Similarly, Aiken et al. \cite{aiken2020neural} also proposed to remove the hidden backdoor by pruning DNNs based on the synthetic trigger from another perspective. Moreover, Shen et al. \cite{shen2021backdoor} proposed an efficient trigger synthesis based defense. Different from previous defenses, which needed to generate all potential triggers towards each class, this defense selects only one class for trigger optimization in each round, inspired by the K-Arm bandit \cite{auer2002finite}. Recently, Hu et al. \cite{hu2022trigger} designed a topological prior to improve the quality of trigger synthesis. Note that all previous defenses are white-box, requiring defenders have the access to model source files. Most recently, a few black-box synthesis-based defenses \cite{dong2021black,guo2021aeva} were also proposed, where defenders can reverse trigger patterns even when they can only obtain model predictions ($e.g.$, probability vectors or predicted labels).

\vspace{0.4em}
\subsubsection{Model Diagnosis based Defenses}
These defenses justify whether a suspicious model is infected based on a pre-trained meta-classifier and refuse to deploy infected models. Since only the benign models are used for deployment, it naturally eliminates the hidden backdoor.

To the best of our knowledge, Kolouri el al. \cite{kolouri2020universal} first discussed how to diagnose a given model. Specifically, they jointly optimized some universal litmus patterns (ULPs) and a meta-classifier, which was further used to diagnose suspicious models based on the predictions of obtained ULPs. Different from the previous defense where both infected models and benign models are required to train the meta-classifier, an effective meta-classiﬁer can be trained only on benign models based on the strategies proposed in \cite{xu2019detecting}. Besides, motivated by the observation that the heatmaps from benign and infected models have different characteristics, Huang et al. \cite{huang2019neuroninspect} adopted an outlier detector as the meta-classifier based on three extracted features of generated saliency maps. In \cite{huangone20}, they designed an one-pixel signature representation, based on which to distinguish benign and infected models. Besides, Wang et al. \cite{wang2020practical} discussed how to detect whether a given mode is benign or infected in the data-limited and data-free cases. Most recently, \cite{zheng2021topological} revealed that benign models and infected DNNs have significant topologically structural differences, which can be used to diagnose suspicious models.

\vspace{0.4em}
\subsubsection{Poison Suppression based Defenses}
\label{sec:PoisonSuppression}
These defenses depress the effectiveness of poisoned samples during the training process to prevent the creation of hidden backdoors. Du et al. \cite{du2020} first explored poison suppression based defenses, where they adopted noisy SGD to learn differentially private DNNs for the defense. With the randomness in the training process, the malicious effects of poisoned samples were reduced by random noise, preventing backdoor creation. Motivated by the observation that the $\ell^2$ norm of the gradients of poisoned samples have signiﬁcantly higher magnitudes than those of benign samples and their gradient orientations are also different, Hong et al. \cite{hong2020effectiveness} adopted differentially private stochastic gradient descent (DPSGD) to clip and perturb individual gradients during the training process. Accordingly, the trained model had no hidden backdoor as well as its robustness towards targeted adversarial attacks was also increased. Besides, Borgain et al. \cite{borgnia2021strong} revealed that introducing strong data augmentation methods ($e.g.$, CutMix \cite{devries2017cutout}) can effectively prevent the creation of hidden backdoors for they significantly perturbed the trigger patterns during the training process. Li et al. \cite{li2021anti} proposed a gradient ascent based anti-backdoor method, based on the observations that backdoor attacks have faster learning on poisoned data and target-class dependency. Most recently, Huang et al. \cite{huang2022backdoor} revealed that the hidden backdoors are learned mostly due to the end-to-end supervised training paradigm, based on which they proposed a simple yet effective decoupling-based training method for backdoor suppression.

\vspace{1em}
\subsubsection{Training Sample Filtering based Defenses}
These defenses aim at filtering poisoned samples from the training dataset. After the filtering process, only benign samples or purified poisoned samples will be used in the training process, which eliminates backdoor creation from the source.

To the best of our knowledge, Tran et al. \cite{tran2018spectral} first explored how to filter malicious samples from the training set. Specifically, they demonstrated that poisoned samples tend to leave behind a detectable trace in the spectrum of the covariance of feature representations, which can be used to filter poisoned samples from the training set. Recently, Hayase et al. \cite{hayase2021spectre} introduced robust covariance estimation to amplify the spectral signature of poisoned samples, based on which they designed a more effective filtering method ($i.e.$, SPECTRE). Also inspired by the idea that poisoned samples and benign samples should have different characteristics in the hidden feature space, Chen et al. \cite{chen2019detecting} proposed a two-stage filtering method, including \textbf{(1)} clustering the activations of training samples in each class into two clusters and \textbf{(2)} determining which, if any, of the clusters corresponds to poisoned samples. A similar idea was also explored in \cite{soremekun2020exposing}. However, Tang et al. \cite{tang2019demon} demonstrated that simple target contamination can cause the representation of poisoned samples to be less distinguishable from that of benign ones, therefore most of existing filtering-based defenses can be easily bypassed. To address this problem, they proposed a more robust sample filter, based on representation decomposition and its statistical analysis. Different from previous methods, Chan et al. \cite{chan2019poison} separated poisoned samples based on signals contained in input gradients. A similar idea was explored in \cite{chou2020sentinet}, where they adopted the saliency map to identify trigger regions and filter poisoned samples. Besides, Wang et al. \cite{wang2021unified} formulated the filtering as optimal data selection, based on which they proposed a unified framework to filter different types of malicious training samples. Most recently, Zeng et al. \cite{zeng2021rethinking} revealed that poisoned samples of existing attacks had some high-frequency artifacts even if their trigger patterns are invisible in the input space. Based on this observation, they designed a simple yet effective filtering method based on those artifacts.

\vspace{1em}
\subsubsection{Testing Sample Filtering based Defenses}
These defenses also filter malicious samples, whereas the filtering happened in the inference instead of the training process. Only benign testing or purified attacked samples will be predicted by the deployed model. These defenses prevent backdoor activation for they can remove trigger patterns.

Motivated by the observation that most of the existing backdoor triggers are input-agnostic, Gao et al. \cite{gao2019strip} proposed to filter attacked samples via superimposing various image patterns on the suspicious samples. The smaller the randomness among the predictions of perturbed inputs, the higher the probability that the suspicious sample is attacked. In \cite{subedar2019deep}, Subedar et al. adopted model uncertainty to distinguish between benign and attacked samples. After that, Du et al. \cite{du2020} treated the filtering as outlier detection, based on which they proposed a differential privacy based filtering method. Besides, Jin et al. \cite{jin2020unified} proposed to detect attacked samples based on existing detection-based adversarial defenses \cite{feinman2017detecting,ma2018characterizing,wang2019adversarial}. Most recently, a lightweight method was proposed in \cite{javaheripi2020cleann}, which can filter attacked samples without labeled samples or prior assumptions on trigger patterns.

\begin{table*}[ht]
\centering
\vspace{-1em}
\caption{Summary of benchmark datasets used in image recognition.}
\vspace{-0.5em}
\label{tab:datasets}
\scalebox{0.97}{
\begin{tabular}{c|c|ccc|l}
\toprule
Category                                   & Datasets          & \# Image Size & \# Training Samples & \# Testing Samples & Cited Literature \\ \hline
\multirow{14}{*}{Natural Image Recognition}
& MNIST \cite{MNIST} & $28\times28$ & 60,000 & 10,000 & \tabincell{c}{\cite{liu2017neural,dumford2018backdooring,huang2019neuroninspect,xu2019detecting, udeshi2019model}\\ \cite{wang2019neural,chen2019detecting,gu2019badnets,chen2019deepinspect,gao2019strip}\\ \cite{subedar2019deep,zhong2020backdoor,umer2020targeted,liu2020backdoor,sommer2020towards}\\ \cite{aiken2020neural,kolouri2020universal,du2020,wang2020certifying,weber2020rab}\\ \cite{soremekun2020exposing,jin2020unified,shan2020using,zhu2020gangsweep,gao2020analyzing}\\
\cite{nguyen2020wanet,nguyen2020input,huangone20,javaheripi2020cleann,yoshida2020disabling}\\
\cite{doan2021lira,doan2021backdoor,wang2021backdoor,liu2021removing,xiang2022post}} \\ \cline{2-6} 
& Fashion MNIST \cite{fashionMNIST} & $28\times28$ & 60,000 & 10,000 & \cite{huangone20,hong2020effectiveness,soremekun2020exposing,xiang2022post}\\ \cline{2-6} 
& CIFAR \cite{CIFAR} & $32\times32\times3$ & 50,000 & 10,000 & \tabincell{c}{\cite{adi2018turning,tran2018spectral,turner2019label,li2020invisible,doan2019februus}\\ \cite{qiao2019defending,xu2019detecting,gao2019strip,subedar2019deep,chan2019poison}\\ \cite{zhong2020backdoor,saha2020hidden,garg2020can,quiring2020backdooring,sommer2020towards}\\ \cite{aiken2020neural,kolouri2020universal,veldanda2020nnoculation,hong2020effectiveness,weber2020rab}\\ \cite{gao2020analyzing,soremekun2020exposing, guo2020trojannet, rakin2020tbt,li2021backdoor}\\ \cite{zhu2020gangsweep, harikumar2020scalable, wang2020practical,shan2020using,zhao2020bridging}\\
\cite{nguyen2020wanet,nguyen2020input,li2020open,lin2020composite,xiang2020detection}\\ \cite{zeng2020deepsweep,cheng2021deep,li2021neural,doan2021lira,doan2021backdoor}\\
\cite{hammoud2021check,wang2021backdoor,souri2021sleeper,shumailov2021manipulating,zhang2021poison}\\
\cite{chen2021proflip,dong2021black,li2022defending,zhao2021deep,zhang2021inject}\\
\cite{qi2021subnet,wu2021adversarial,guo2021aeva,borgnia2021strong,liu2021removing}\\
\cite{li2021anti,hayase2021spectre,wang2021unified,zeng2021rethinking,xiang2022post}} \\ \cline{2-6} 
& SVHN \cite{SVHN} & $32\times32\times3$ & 73,257 & 26,032 & \cite{zhao2020bridging,guo2020trojannet, rakin2020tbt,chen2021proflip} \\ \cline{2-6} 
& ImageNet \cite{ImageNet} & $224\times224\times3$ & 1,281,167 & 50,000 & \tabincell{c}{ \cite{adi2018turning,guo2019tabor,tang2019demon,bagdasaryan2020blind,liu2020reflection}\\ \cite{saha2020hidden,chou2020sentinet,weber2020rab,gao2020analyzing,tang2020embarrassingly}\\ \cite{rakin2020tbt,zhu2020gangsweep,huangone20,wang2020practical,cheng2021deep}\\ \cite{li2021invisible,shen2021backdoor,doan2021lira,doan2021backdoor,wang2021backdoor}\\
\cite{souri2021sleeper,zhang2021poison,chen2021proflip,dong2021black,wenger2020backdoor}\\
\cite{li2022defending,lin2020you,zhao2021deep,zhang2021inject,li2021deeppayload}\\
\cite{guo2021aeva,li2021anti,xiang2022post}} \\ \hline
\multirow{2}{*}{Traffic Sign Recognition}  
& GTSRB \cite{GTSRB} & --- & 34,799 & 12,630 & \tabincell{c}{\cite{li2020invisible,doan2019februus,chen2019deepinspect,cheng2019defending,huang2019neuroninspect}\\ \cite{wang2019neural,guo2019tabor,gao2019strip,tang2019demon,zhong2020backdoor}\\ \cite{liu2020reflection,villarreal2020confoc,kolouri2020universal,veldanda2020nnoculation,gao2020analyzing}\\ \cite{guo2020trojannet,tang2020embarrassingly,harikumar2020scalable,jin2020unified,shan2020using}\\ \cite{li2020open,zhu2020gangsweep,nguyen2020input, huangone20,wang2020practical}\\ \cite{yoshida2020disabling,zeng2020deepsweep,javaheripi2020cleann,nguyen2020wanet,cheng2021deep}\\
\cite{li2021neural,doan2021lira,doan2021backdoor,hammoud2021check,wang2021backdoor}\\
\cite{dong2021black,zhang2021poison,li2021anti,zeng2021rethinking}} \\  \cline{2-6} 
& U.S. Traffic Sign \cite{USTS} & --- & 6,889 & 1,724            & \cite{liu2018fine,gu2019badnets,udeshi2019model} \\  \hline
\multirow{8}{*}{Face Recognition} 
& YouTube Face \cite{YouTubeFace} & --- & \multicolumn{2}{c|}{3,425 videos of 1,595 people}           & \tabincell{c}{ \cite{chen2017targeted,liu2018fine,wang2019neural,veldanda2020nnoculation,tang2020embarrassingly}\\ \cite{shan2020using,lin2020composite}} \\ \cline{2-6} 
& PubFig \cite{PubFig} & --- & \multicolumn{2}{c|}{58,797 images of 200 people} & \tabincell{c}{\cite{wang2019neural,liu2020reflection,tang2020embarrassingly,jin2020unified,zeng2020deepsweep}\\
\cite{wang2021backdoor,wang2021unified,zeng2021rethinking}} \\  \cline{2-6} 
& VGGFace \cite{BMVC} & --- & \multicolumn{2}{c|}{2.6 million images of 2,622 people} & \tabincell{c}{ \cite{liutrojaning,udeshi2019model,chen2019deepinspect,wang2019neural,villarreal2020confoc}\\ \cite{wang2020backdoor,chou2020sentinet,zhu2020gangsweep,javaheripi2020cleann,cheng2021deep}\\
\cite{zhang2021poison,wenger2020backdoor,qi2021subnet}} \\ \cline{2-6} 
& VGGFace2 \cite{vggface2} & --- & \multicolumn{2}{c|}{3.3 million images of 9,131 people}           & \cite{dumford2018backdooring,doan2019februus,wenger2020backdoor} \\ \cline{2-6} 
& LFW \cite{LFW} & --- & \multicolumn{2}{c|}{13,233 images of 5,749 people} & \cite{guo2019tabor,wang2020backdoor,chou2020sentinet} \\ \bottomrule
\end{tabular}
}
\footnotesize{\textbf{Note}: \textbf{(1)} The sign sizes vary from $6\times6$ to $167\times168$ pixels in the U.S. Traffic Sign dataset; \textbf{(2)} There is no given division between the training set and the testing set in most face recognition datasets. Users need to divide the dataset by themselves according to their needs.}
\end{table*}

\vspace{1em}
\subsection{Certified Backdoor Defenses}\label{sec: certified}
Although multiple empirical defenses have been proposed and reached decent performance against some backdoor attacks, almost all of them were bypassed by following adaptive attacks \cite{Tan2020Bypassing,Aniruddha2020Hidden}. To terminate this `cat-and-mouse chasing game', Wang et al. \cite{wang2020certifying} took the ﬁrst step towards the certiﬁed defense against backdoor attacks based on the \emph{random smoothing} technique \cite{cohen2019certified}. Randomized smoothing was originally developed to certify robustness against adversarial examples, where the smoothed function was built from the base function via adding random noise to the data vector to certify the robustness of a classiﬁer under certain conditions. Similar to \cite{Rosenfeld2020Certified}, Wang et al. treated the entire training procedure of the classiﬁer as the base function to generalize classical randomized smoothing to defend against backdoor attacks. In \cite{weber2020rab}, Weber et al. demonstrated that directly applying randomized smoothing, as in \cite{wang2020certifying}, will not provide high certiﬁed robustness bounds. Instead, they proposed a uniﬁed framework with the examination of different smoothing noise distributions and provided a tightness analysis for the robustness bound. Most recently, a few studies \cite{jia2021intrinsic,levine2021deep,jia2022certified} also adopted ensemble techniques ($e.g.$, Bagging \cite{breiman1996bagging}) in designing certiﬁed defenses to further improve effectiveness.

\subsection{Evaluation Metrics}

\noindent \textbf{Metrics for Detection-like Empirical Defenses. } 
Model diagnosis based defenses and testing sample filtering based defenses are all detection-like methods, whose main target is to identify whether a suspicious object ($e.g.$, a trained DNN or sample) is malicious. This is essentially a binary classification problem. To evaluate their performance, three metrics \cite{sugiyama2015introduction}, including \textbf{(1)} precision, \textbf{(2)} recall, and \textbf{(3)} F1-score, are usually adopted. The higher the precision, recall, and F1-score, the better the defense performance.

\vspace{0.4em}
\noindent \textbf{Metrics for Non-detection-like Empirical Defenses. } 
Except for model diagnosis based and testing sample filtering based defenses, all other methods are non-detection-like. Their main target is to achieve correct predictions for both benign and attacked samples. Accordingly, both benign accuracy and attack success rate (as defined in Section \ref{Sec:def}) are also adopted for the evaluation. In particular, although a detection process is also involved in training sample filtering based defenses, three metrics ($i.e.$, precision, recall, and F1-score) described above are not suitable for their evaluation. These defenses may try to discard as many poisoned samples as possible to reduce the possibility of creating hidden backdoors trained on the filtered dataset, even with the sacrifice of certain benign samples.

\vspace{0.4em}
\noindent \textbf{Metrics for Certified Defenses. } 
As mentioned in Section \ref{sec: certified}, existing certified backdoor defenses all adopted random smoothing. As such, these methods can provide a certified radius, where all perturbation within the $\ell^p$ ball with the certified radius can not change the prediction of the model under certain assumptions.
To evaluate their performance, people usually use the \textbf{(1)} benign accuracy, \textbf{(2)} certified rate, and \textbf{(3)} certified accuracy as the evaluation metric \cite{wang2020certifying,weber2020rab}. Specifically, the benign accuracy indicates how well the (smoothed) classifier performs in classifying benign samples; the certified rate is the fraction of samples that can be certified at radius greater than the certified radius; and the certified accuracy is the fraction of the test set which is classified correctly and is certified as robust with a radius greater than the certified radius. The greater the benign accuracy, certified rate, and certified accuracy, the better the defense performance.

\vspace{1em}
\section{Benchmark Datasets}\label{sec:dataset}
Similar to that of adversarial learning and data poisoning, most of the existing backdoor-related literature focused on image classification tasks. In this section, we summarize all classical image classification benchmark datasets in Table \ref{tab:datasets}. 

Specifically, these benchmark datasets can be divided into three main categories, including \emph{natural image recognition}, \emph{traffic sign recognition}, and \emph{face recognition}. The former ones are classical in the image classification, while the second and third ones are tasks requiring strict security guarantees. We recommend that future studies should be evaluated on these datasets to facilitate comparison and ensure fairness.

\section{Outlook of Future Directions}\label{sec:outlook}


As presented above, many works have been proposed in the literature of backdoor learning, covering several branches and different scenarios. 
However, we believe that the development of this field is still in its infancy, as many critical problems of backdoor learning have not been well studied. 
In this section, we present five potential research directions to inspire the future development of backdoor learning. 

\subsection{Trigger Design}
The effectiveness and efficiency of poisoning-based backdoor attacks are closely related to their trigger patterns. However, the trigger of most existing attacks was designed in a heuristic ($e.g.$, design with universal perturbation) or even a non-optimized way. How to better optimize the trigger pattern is still an important open question. Besides, only the effectiveness and trigger invisibility were considered in the trigger design. Other criteria, such as minimal poisoning rate and trigger generalization, are also worth further exploration.

\subsection{Semantic and Physical Backdoor Attacks}
As presented in Section \ref{sec: attacks for image and video recognition}, semantic and physical attacks are more serious threats to AI systems in practical scenarios, while their studies are left far behind compared to other types of backdoor attacks. 
More thorough studies to obtain better understandings of these attacks would be important steps towards alleviating the backdoor threats in practice.

\subsection{Attacks Towards Other Tasks}
The success of backdoor attacks heavily relied on the specific trigger design according to the characteristics of the target task. For example, the visual invisibility of the trigger is one of the critical criteria in visual tasks, which ensures stealthiness. 
However, the design of backdoor triggers in different tasks could be quite different ($e.g.$, hiding a trigger into a sentence when attacking NLP-related tasks is quite different from hiding a trigger into an image).
Accordingly, it is necessary to study task-specified backdoor attacks. Currently, existing backdoor attacks mainly focused on computer vision tasks, especially image classification. The research towards other tasks (\eg, recommendation system, speech recognition, and natural language processing) have not been well studied.

\subsection{Effective and Efficient Defenses}
Although many types of empirical backdoor defenses have been proposed (as demonstrated in Section \ref{sec:defense}), almost all of them can be bypassed by subsequent adaptive attacks. 
Besides, except for the preprocessing-based defenses, existing defenses usually suffer from high computational costs. More efforts on designing effective and efficient defenses ($e.g.$, analyzing the weaknesses of existing attacks and how to reduce the computational costs of defenses) should be made to keep up the fast pace of backdoor attacks. Besides, how to design black-box defenses is also worth more attention since these methods are more practical in reality. Moreover, certified backdoor defenses are important yet currently have been rarely studied, which deserve more explorations.

\subsection{Mechanism Exploration}
The principle of backdoor generation and the activation mechanism of backdoor triggers are the holy grail problems in backdoor learning. For example, why hidden backdoors can be created and what happens inside the infected models when the trigger appears have not been carefully studied in existing works. A deeper understanding of the intrinsic mechanism of backdoor attacks can guide the design of more effective attacks and defenses, and the understandings of DNN's behaviors.

\vspace{1em}
\section{Conclusion}\label{sec:conclusion}
Backdoor learning, including backdoor attacks and backdoor defenses, is a critical and booming research area. In this survey, we summarized and categorized existing backdoor attacks and proposed a unified framework for analyzing poisoning-based backdoor attacks. We also discussed the relation between backdoor attacks and related research areas and analyzed existing defenses. Classical benchmark datasets and potential research directions were illustrated at the end. Note that almost all studies in this field were completed in the last four years and the cat-and-mouse game between attacks and defenses is likely to continue in the future. We hope that this survey could remind researchers of backdoor threats and provide a timely view. It would be an important step towards building more robust and safer deep learning methods.

\vspace{1em}
\section*{Acknowledgements}
This work was partly done when Yiming Li was a research intern at Tencent AI Lab, supported by the Tencent Rhino-Bird  Elite Training Program (2020). This work is also supported in part by the National Natural Science Foundation of China under grant 62171248, the Guangdong Province Key Area R\&D Program under Grant 2018B010113001, the R\&D Program of Shenzhen (JCYJ20180508152204044), the PCNL KEY project (PCL2021A07), and the Shenzhen Philosophical and Social Science Plan (SZ2020D009). We also sincerely thank Dr. Baoyuan Wu from the Chinese University of Hong Kong (Shenzhen) and Dr. Bo Li from the University of Illinois Urbana-Champaign for their helpful comments on an early draft of this paper.

\bibliographystyle{IEEEtran}
\bibliography{ref}

\begin{thebibliography}{100}
\providecommand{\url}[1]{#1}
\csname url@samestyle\endcsname
\providecommand{\newblock}{\relax}
\providecommand{\bibinfo}[2]{#2}
\providecommand{\BIBentrySTDinterwordspacing}{\spaceskip=0pt\relax}
\providecommand{\BIBentryALTinterwordstretchfactor}{4}
\providecommand{\BIBentryALTinterwordspacing}{\spaceskip=\fontdimen2\font plus
\BIBentryALTinterwordstretchfactor\fontdimen3\font minus
  \fontdimen4\font\relax}
\providecommand{\BIBforeignlanguage}[2]{{%
\expandafter\ifx\csname l@#1\endcsname\relax
\typeout{** WARNING: IEEEtran.bst: No hyphenation pattern has been}%
\typeout{** loaded for the language `#1'. Using the pattern for}%
\typeout{** the default language instead.}%
\else
\language=\csname l@#1\endcsname
\fi
#2}}
\providecommand{\BIBdecl}{\relax}
\BIBdecl

\bibitem{goodfellow2014explaining}
I.~J. Goodfellow, J.~Shlens, and C.~Szegedy, ``Explaining and harnessing
  adversarial examples,'' in \emph{ICLR}, 2015.

\bibitem{madry2018towards}
A.~Madry, A.~Makelov, L.~Schmidt, D.~Tsipras, and A.~Vladu, ``Towards deep
  learning models resistant to adversarial attacks,'' in \emph{ICLR}, 2018.

\bibitem{bai2020targeted}
J.~Bai, B.~Chen, Y.~Li, D.~Wu, W.~Guo, S.-t. Xia, and E.-h. Yang, ``Targeted
  attack for deep hashing based retrieval,'' in \emph{ECCV}, 2020.

\bibitem{wu2020adversarial}
D.~Wu, S.~Xia, and Y.~Wang, ``Adversarial weight perturbation helps robust
  generalization,'' in \emph{NeurIPS}, 2020.

\bibitem{bai2021improving}
Y.~Bai, Y.~Zeng, Y.~Jiang, S.-T. Xia, X.~Ma, and Y.~Wang, ``Improving
  adversarial robustness via channel-wise activation suppressing,'' in
  \emph{ICLR}, 2021.

\bibitem{li2022semi}
Y.~Li, B.~Wu, Y.~Feng, Y.~Fan, Y.~Jiang, Z.~Li, and S.-T. Xia,
  ``Semi-supervised robust training with generalized perturbed neighborhood,''
  \emph{Pattern Recognition}, vol. 124, p. 108472, 2022.

\bibitem{gu2019badnets}
T.~Gu, K.~Liu, B.~Dolan-Gavitt, and S.~Garg, ``Badnets: Evaluating backdooring
  attacks on deep neural networks,'' \emph{IEEE Access}, vol.~7, pp.
  47\,230--47\,244, 2019.

\bibitem{liu2020reflection}
Y.~Liu, X.~Ma, J.~Bailey, and F.~Lu, ``Reflection backdoor: A natural backdoor
  attack on deep neural networks,'' in \emph{ECCV}, 2020.

\bibitem{li2021backdoor}
Y.~Li, T.~Zhai, Y.~Jiang, Z.~Li, and S.-T. Xia, ``Backdoor attack in the
  physical world,'' in \emph{ICLR Workshop}, 2021.

\bibitem{chen2017targeted}
X.~Chen, C.~Liu, B.~Li, K.~Lu, and D.~Song, ``Targeted backdoor attacks on deep
  learning systems using data poisoning,'' \emph{arXiv preprint
  arXiv:1712.05526}, 2017.

\bibitem{li2020invisible}
S.~Li, M.~Xue, B.~Zhao, H.~Zhu, and X.~Zhang, ``Invisible backdoor attacks on
  deep neural networks via steganography and regularization,'' \emph{IEEE
  Transactions on Dependable and Secure Computing}, 2020.

\bibitem{li2021invisible}
Y.~Li, Y.~Li, B.~Wu, L.~Li, R.~He, and S.~Lyu, ``Invisible backdoor attack with
  sample-specific triggers,'' in \emph{ICCV}, 2021.

\bibitem{turner2019label}
A.~Turner, D.~Tsipras, and A.~Madry, ``Label-consistent backdoor attacks,''
  \emph{arXiv preprint arXiv:1912.02771}, 2019.

\bibitem{saha2020hidden}
A.~Saha, A.~Subramanya, and H.~Pirsiavash, ``Hidden trigger backdoor attacks,''
  in \emph{AAAI}, 2020.

\bibitem{zhao2020clean}
S.~Zhao, X.~Ma, X.~Zheng, J.~Bailey, J.~Chen, and Y.-G. Jiang, ``Clean-label
  backdoor attacks on video recognition models,'' in \emph{CVPR}, 2020.

\bibitem{kurita2020weight}
K.~Kurita, P.~Michel, and G.~Neubig, ``Weight poisoning attacks on pre-trained
  models,'' in \emph{ACL}, 2020.

\bibitem{wang2020backdoor}
S.~Wang, S.~Nepal, C.~Rudolph, M.~Grobler, S.~Chen, and T.~Chen, ``Backdoor
  attacks against transfer learning with pre-trained deep learning models,''
  \emph{IEEE Transactions on Services Computing}, 2020.

\bibitem{ge2021anti}
Y.~Ge, Q.~Wang, B.~Zheng, X.~Zhuang, Q.~Li, C.~Shen, and C.~Wang,
  ``Anti-distillation backdoor attacks: Backdoors can really survive in
  knowledge distillation,'' in \emph{ACM MM}, 2021.

\bibitem{dumford2018backdooring}
J.~Dumford and W.~Scheirer, ``Backdooring convolutional neural networks via
  targeted weight perturbations,'' \emph{arXiv preprint arXiv:1812.03128},
  2018.

\bibitem{rakin2020tbt}
A.~S. Rakin, Z.~He, and D.~Fan, ``Tbt: Targeted neural network attack with bit
  trojan,'' in \emph{CVPR}, 2020.

\bibitem{chen2021proflip}
H.~Chen, C.~Fu, J.~Zhao, and F.~Koushanfar, ``Proflip: Targeted trojan attack
  with progressive bit flips,'' in \emph{ICCV}, 2021.

\bibitem{tang2020embarrassingly}
R.~Tang, M.~Du, N.~Liu, F.~Yang, and X.~Hu, ``An embarrassingly simple approach
  for trojan attack in deep neural networks,'' in \emph{KDD}, 2020.

\bibitem{li2021deeppayload}
Y.~Li, J.~Hua, H.~Wang, C.~Chen, and Y.~Liu, ``Deeppayload: Black-box backdoor
  attack on deep learning models through neural payload injection,'' in
  \emph{ICSE}, 2021.

\bibitem{qi2021subnet}
X.~Qi, J.~Zhu, C.~Xie, and Y.~Yang, ``Subnet replacement: Deployment-stage
  backdoor attack against deep neural networks in gray-box setting,'' in
  \emph{ICLR Workshop}, 2021.

\bibitem{wang2019neural}
B.~Wang, Y.~Yao, S.~Shan, H.~Li, B.~Viswanath, H.~Zheng, and B.~Y. Zhao,
  ``Neural cleanse: Identifying and mitigating backdoor attacks in neural
  networks,'' in \emph{IEEE S\&P}, 2019.

\bibitem{kolouri2020universal}
S.~Kolouri, A.~Saha, H.~Pirsiavash, and H.~Hoffmann, ``Universal litmus
  patterns: Revealing backdoor attacks in cnns,'' in \emph{CVPR}, 2020.

\bibitem{li2021anti}
Y.~Li, X.~Lyu, N.~Koren, L.~Lyu, B.~Li, and X.~Ma, ``Anti-backdoor learning:
  Training clean models on poisoned data,'' in \emph{NeurIPS}, 2021.

\bibitem{wang2020certifying}
B.~Wang, X.~Cao, N.~Z. Gong \emph{et~al.}, ``On certifying robustness against
  backdoor attacks via randomized smoothing,'' in \emph{CVPR Workshop}, 2020.

\bibitem{weber2020rab}
M.~Weber, X.~Xu, B.~Karlas, C.~Zhang, and B.~Li, ``Rab: Provable robustness
  against backdoor attacks,'' \emph{arXiv preprint arXiv:2003.08904}, 2020.

\bibitem{xie2021crfl}
C.~Xie, M.~Chen, P.-Y. Chen, and B.~Li, ``Crfl: Certifiably robust federated
  learning against backdoor attacks,'' in \emph{ICML}, 2021.

\bibitem{liu2020survey}
Y.~Liu, A.~Mondal, A.~Chakraborty, M.~Zuzak, N.~Jacobsen, D.~Xing, and
  A.~Srivastava, ``A survey on neural trojans,'' in \emph{ISQED}, 2020.

\bibitem{goldblum2020dataset}
M.~Goldblum, D.~Tsipras, C.~Xie, X.~Chen, A.~Schwarzschild, D.~Song, A.~Madry,
  B.~Li, and T.~Goldstein, ``Dataset security for machine learning: Data
  poisoning, backdoor attacks, and defenses,'' \emph{arXiv preprint
  arXiv:2012.10544}, 2020.

\bibitem{kaviani2021defense}
S.~Kaviani and I.~Sohn, ``Defense against neural trojan attacks: A survey,''
  \emph{Neurocomputing}, vol. 423, pp. 651--667, 2021.

\bibitem{gao2020backdoor}
Y.~Gao, B.~G. Doan, Z.~Zhang, S.~Ma, J.~Zhang, A.~Fu, S.~Nepal, and H.~Kim,
  ``Backdoor attacks and countermeasures on deep learning: A comprehensive
  review,'' \emph{arXiv preprint arXiv:2007.10760}, 2020.

\bibitem{li2020deep}
S.~Li, S.~Ma, M.~Xue, and B.~Z.~H. Zhao, ``Deep learning backdoors,''
  \emph{arXiv preprint arXiv:2007.08273}, 2020.

\bibitem{guo2021overview}
W.~Guo, B.~Tondi, and M.~Barni, ``An overview of backdoor attacks against deep
  neural networks and possible defences,'' \emph{arXiv preprint
  arXiv:2111.08429}, 2021.

\bibitem{zhong2020backdoor}
H.~Zhong, C.~Liao, A.~C. Squicciarini, S.~Zhu, and D.~Miller, ``Backdoor
  embedding in convolutional neural network models via invisible
  perturbation,'' in \emph{ACM CODASPY}, 2020.

\bibitem{moosavi2017universal}
S.-M. Moosavi-Dezfooli, A.~Fawzi, O.~Fawzi, and P.~Frossard, ``Universal
  adversarial perturbations,'' in \emph{CVPR}, 2017.

\bibitem{doan2021lira}
K.~Doan, Y.~Lao, W.~Zhao, and P.~Li, ``Lira: Learnable, imperceptible and
  robust backdoor attacks,'' in \emph{ICCV}, 2021.

\bibitem{doan2021backdoor}
K.~Doan, Y.~Lao, and P.~Li, ``Backdoor attack with imperceptible input and
  latent modification,'' in \emph{NeurIPS}, 2021.

\bibitem{nguyen2020wanet}
T.~A. Nguyen and A.~T. Tran, ``Wanet-imperceptible warping-based backdoor
  attack,'' in \emph{ICLR}, 2021.

\bibitem{bagdasaryan2020blind}
E.~Bagdasaryan and V.~Shmatikov, ``Blind backdoors in deep learning models,''
  in \emph{USENIX Security}, 2021.

\bibitem{cheng2021deep}
S.~Cheng, Y.~Liu, S.~Ma, and X.~Zhang, ``Deep feature space trojan attack of
  neural networks by controlled detoxification,'' in \emph{AAAI}, 2021.

\bibitem{hammoud2021check}
H.~A. A.~K. Hammoud and B.~Ghanem, ``Check your other door! establishing
  backdoor attacks in the frequency domain,'' \emph{arXiv preprint
  arXiv:2109.05507}, 2021.

\bibitem{wang2021backdoor}
T.~Wang, Y.~Yao, F.~Xu, S.~An, and T.~Wang, ``Backdoor attack through frequency
  domain,'' \emph{arXiv preprint arXiv:2111.10991}, 2021.

\bibitem{quiring2020backdooring}
E.~Quiring and K.~Rieck, ``Backdooring and poisoning neural networks with
  image-scaling attacks,'' in \emph{IEEE S\&P Workshop}, 2020.

\bibitem{xiao2019seeing}
Q.~Xiao, Y.~Chen, C.~Shen, Y.~Chen, and K.~Li, ``Seeing is not believing:
  Camouflage attacks on image scaling algorithms,'' in \emph{USENIX Security},
  2019.

\bibitem{souri2021sleeper}
H.~Souri, M.~Goldblum, L.~Fowl, R.~Chellappa, and T.~Goldstein, ``Sleeper
  agent: Scalable hidden trigger backdoors for neural networks trained from
  scratch,'' \emph{arXiv preprint arXiv:2106.08970}, 2021.

\bibitem{liu2021investigating}
R.~Liu, J.~Gao, J.~Zhang, D.~Meng, and Z.~Lin, ``Investigating bi-level
  optimization for learning and vision from a unified perspective: A survey and
  beyond,'' \emph{arXiv preprint arXiv:2101.11517}, 2021.

\bibitem{shumailov2021manipulating}
I.~Shumailov, Z.~Shumaylov, D.~Kazhdan, Y.~Zhao, N.~Papernot, M.~A. Erdogdu,
  and R.~Anderson, ``Manipulating sgd with data ordering attacks,'' in
  \emph{NeurIPS}, 2021.

\bibitem{liutrojaning}
Y.~Liu, S.~Ma, Y.~Aafer, W.-C. Lee, J.~Zhai, W.~Wang, and X.~Zhang, ``Trojaning
  attack on neural networks,'' in \emph{NDSS}, 2017.

\bibitem{garg2020can}
S.~Garg, A.~Kumar, V.~Goel, and Y.~Liang, ``Can adversarial weight
  perturbations inject neural backdoors?'' in \emph{CIKM}, 2020.

\bibitem{geiping2020witches}
J.~Geiping, L.~H. Fowl, W.~R. Huang, W.~Czaja, G.~Taylor, M.~Moeller, and
  T.~Goldstein, ``Witches' brew: Industrial scale data poisoning via gradient
  matching,'' in \emph{ICLR}, 2020.

\bibitem{bagdasaryan2020backdoor}
E.~Bagdasaryan, A.~Veit, Y.~Hua, D.~Estrin, and V.~Shmatikov, ``How to backdoor
  federated learning,'' in \emph{AISTATS}, 2020.

\bibitem{lin2020composite}
J.~Lin, L.~Xu, Y.~Liu, and X.~Zhang, ``Composite backdoor attack for deep
  neural network by mixing existing benign features,'' in \emph{CCS}, 2020.

\bibitem{chen2019deepinspect}
H.~Chen, C.~Fu, J.~Zhao, and F.~Koushanfar, ``Deepinspect: A black-box trojan
  detection and mitigation framework for deep neural networks.'' in
  \emph{IJCAI}, 2019.

\bibitem{guo2019tabor}
W.~Guo, L.~Wang, Y.~Xu, X.~Xing, M.~Du, and D.~Song, ``Towards inspecting and
  eliminating trojan backdoors in deep neural networks,'' in \emph{ICDM}, 2020.

\bibitem{shen2021backdoor}
G.~Shen, Y.~Liu, G.~Tao, S.~An, Q.~Xu, S.~Cheng, S.~Ma, and X.~Zhang,
  ``Backdoor scanning for deep neural networks through k-arm optimization,''
  \emph{arXiv preprint arXiv:2102.05123}, 2021.

\bibitem{dong2021black}
Y.~Dong, X.~Yang, Z.~Deng, T.~Pang, Z.~Xiao, H.~Su, and J.~Zhu, ``Black-box
  detection of backdoor attacks with limited information and data,'' in
  \emph{ICCV}, 2021.

\bibitem{huang2019neuroninspect}
X.~Huang, M.~Alzantot, and M.~Srivastava, ``Neuroninspect: Detecting backdoors
  in neural networks via output explanations,'' \emph{arXiv preprint
  arXiv:1911.07399}, 2019.

\bibitem{chou2020sentinet}
E.~Chou, F.~Tram{\`e}r, and G.~Pellegrino, ``Sentinet: Detecting localized
  universal attacks against deep learning systems,'' in \emph{IEEE S\&P
  Workshop}, 2020.

\bibitem{nguyen2020input}
A.~Nguyen and A.~Tran, ``Input-aware dynamic backdoor attack,'' in
  \emph{NeurIPS}, 2020.

\bibitem{zhang2021poison}
J.~Zhang, D.~Chen, J.~Liao, Q.~Huang, G.~Hua, W.~Zhang, and N.~Yu, ``Poison
  ink: Robust and invisible backdoor attack,'' \emph{arXiv preprint
  arXiv:2108.02488}, 2021.

\bibitem{wenger2020backdoor}
E.~Wenger, J.~Passananti, A.~N. Bhagoji, Y.~Yao, H.~Zheng, and B.~Y. Zhao,
  ``Backdoor attacks against deep learning systems in the physical world,'' in
  \emph{CVPR}, 2021.

\bibitem{dai2019backdoor}
J.~Dai, C.~Chen, and Y.~Li, ``A backdoor attack against lstm-based text
  classification systems,'' \emph{IEEE Access}, vol.~7, pp. 138\,872--138\,878,
  2019.

\bibitem{chen2020badnl}
X.~Chen, A.~Salem, D.~Chen, M.~Backes, S.~Ma, Q.~Shen, Z.~Wu, and Y.~Zhang,
  ``Badnl: Backdoor attacks against nlp models with semantic-preserving
  improvements,'' in \emph{ACSAC}, 2021.

\bibitem{qi2021turn}
F.~Qi, Y.~Yao, S.~Xu, Z.~Liu, and M.~Sun, ``Turn the combination lock:
  Learnable textual backdoor attacks via word substitution,'' in \emph{ACL},
  2021.

\bibitem{qi2021hidden}
F.~Qi, M.~Li, Y.~Chen, Z.~Zhang, Z.~Liu, Y.~Wang, and M.~Sun, ``Hidden killer:
  Invisible textual backdoor attacks with syntactic trigger,'' in \emph{ACL},
  2021.

\bibitem{yang2021rethinking}
W.~Yang, Y.~Lin, P.~Li, J.~Zhou, and X.~Sun, ``Rethinking stealthiness of
  backdoor attack against nlp models,'' in \emph{ACL}, 2021.

\bibitem{qi2021mind}
F.~Qi, Y.~Chen, X.~Zhang, M.~Li, Z.~Liu, and M.~Sun, ``Mind the style of text!
  adversarial and backdoor attacks based on text style transfer,'' in
  \emph{EMNLP}, 2021.

\bibitem{yang2021careful}
W.~Yang, L.~Li, Z.~Zhang, X.~Ren, X.~Sun, and B.~He, ``Be careful about
  poisoned word embeddings: Exploring the vulnerability of the embedding layers
  in nlp models,'' in \emph{NAACL}, 2021.

\bibitem{li2021backdoora}
L.~Li, D.~Song, X.~Li, J.~Zeng, R.~Ma, and X.~Qiu, ``Backdoor attacks on
  pre-trained models by layerwise weight poisoning,'' in \emph{EMNLP}, 2021.

\bibitem{zhang2020backdoor}
Z.~Zhang, J.~Jia, B.~Wang, and N.~Z. Gong, ``Backdoor attacks to graph neural
  networks,'' in \emph{NeurIPS Workshop}, 2020.

\bibitem{xi2020graph}
Z.~Xi, R.~Pang, S.~Ji, and T.~Wang, ``Graph backdoor,'' in \emph{USENIX
  Security}, 2021.

\bibitem{chen2021dyn}
J.~Chen, H.~Xiong, H.~Zheng, J.~Zhang, G.~Jiang, and Y.~Liu, ``Dyn-backdoor:
  Backdoor attack on dynamic link prediction,'' \emph{arXiv preprint
  arXiv:2110.03875}, 2021.

\bibitem{tian2021poisoning}
G.~Tian, W.~Jiang, W.~Liu, and Y.~Mu, ``Poisoning morphnet for clean-label
  backdoor attack to point clouds,'' \emph{arXiv preprint arXiv:2105.04839},
  2021.

\bibitem{xiang2021backdoor}
Z.~Xiang, D.~J. Miller, S.~Chen, X.~Li, and G.~Kesidis, ``A backdoor attack
  against 3d point cloud classifiers,'' in \emph{ICCV}, 2021.

\bibitem{li2021pointba}
X.~Li, Z.~Chen, Y.~Zhao, Z.~Tong, Y.~Zhao, A.~Lim, and J.~T. Zhou, ``Pointba:
  Towards backdoor attacks in 3d point cloud,'' in \emph{ICCV}, 2021.

\bibitem{yan2021deep}
Z.~Yan, J.~Wu, G.~Li, S.~Li, and M.~Guizani, ``Deep neural backdoor in
  semi-supervised learning: Threats and countermeasures,'' \emph{IEEE
  Transactions on Information Forensics and Security}, vol.~16, pp. 4827--4842,
  2021.

\bibitem{jia2022badencoder}
J.~Jia, Y.~Liu, and N.~Z. Gong, ``Badencoder: Backdoor attacks to pre-trained
  encoders in self-supervised learning,'' in \emph{IEEE S\&P}, 2022.

\bibitem{carlini2022poisoning}
N.~Carlini and A.~Terzis, ``Poisoning and backdooring contrastive learning,''
  in \emph{ICLR}, 2022.

\bibitem{kiourti2020trojdrl}
P.~Kiourti, K.~Wardega, S.~Jha, and W.~Li, ``Trojdrl: evaluation of backdoor
  attacks on deep reinforcement learning,'' in \emph{DAC}, 2020.

\bibitem{wang2021backdoorl}
L.~Wang, Z.~Javed, X.~Wu, W.~Guo, X.~Xing, and D.~Song, ``Backdoorl: Backdoor
  attack against competitive reinforcement learning,'' in \emph{IJCAI}, 2021.

\bibitem{ashcraft2021poisoning}
C.~Ashcraft and K.~Karra, ``Poisoning deep reinforcement learning agents with
  in-distribution triggers,'' in \emph{ICLR Workshop}, 2021.

\bibitem{ma2021quantization}
H.~Ma, H.~Qiu, Y.~Gao, Z.~Zhang, A.~Abuadbba, A.~Fu, S.~Al-Sarawi, and
  D.~Abbott, ``Quantization backdoors to deep learning models,'' \emph{arXiv
  preprint arXiv:2108.09187}, 2021.

\bibitem{hong2021qu}
S.~Hong, M.-A. Panaitescu-Liess, Y.~Kaya, and T.~Dumitras, ``Qu-anti-zation:
  Exploiting quantization artifacts for achieving adversarial outcomes,'' in
  \emph{NeurIPS}, 2021.

\bibitem{pan2021understanding}
X.~Pan, M.~Zhang, Y.~Yan, and M.~Yang, ``Understanding the threats of trojaned
  quantized neural network in model supply chains,'' in \emph{ACSAC}, 2021.

\bibitem{zhai2021backdoor}
T.~Zhai, Y.~Li, Z.~Zhang, B.~Wu, Y.~Jiang, and S.-T. Xia, ``Backdoor attack
  against speaker verification,'' in \emph{ICASSP}, 2021.

\bibitem{koffas2021can}
S.~Koffas, J.~Xu, M.~Conti, and S.~Picek, ``Can you hear it? backdoor attacks
  via ultrasonic triggers,'' \emph{arXiv preprint arXiv:2107.14569}, 2021.

\bibitem{li2021backdoormal}
C.~Li, X.~Chen, D.~Wang, S.~Wen, M.~E. Ahmed, S.~Camtepe, and Y.~Xiang,
  ``Backdoor attack on machine learning based android malware detectors,''
  \emph{IEEE Transactions on Dependable and Secure Computing}, 2021.

\bibitem{severi2021explanation}
G.~Severi, J.~Meyer, S.~Coull, and A.~Oprea, ``Explanation-guided backdoor
  poisoning attacks against malware classifiers,'' in \emph{USENIX Security},
  2021.

\bibitem{li2021hidden}
Y.~Li, Y.~Li, Y.~Lv, Y.~Jiang, and S.-T. Xia, ``Hidden backdoor attack against
  semantic segmentation models,'' in \emph{ICLR Workshop}, 2021.

\bibitem{fang2022backdoor}
S.~Fang and A.~Choromanska, ``Backdoor attacks on the dnn interpretation
  system,'' in \emph{AAAI}, 2022.

\bibitem{li2022few}
Y.~Li, H.~Zhong, X.~Ma, Y.~Jiang, and S.-T. Xia, ``Few-shot backdoor attacks on
  visual object tracking,'' in \emph{ICLR}, 2022.

\bibitem{bhagoji2019analyzing}
A.~N. Bhagoji, S.~Chakraborty, P.~Mittal, and S.~Calo, ``Analyzing federated
  learning through an adversarial lens,'' in \emph{ICML}, 2019.

\bibitem{xie2019dba}
C.~Xie, K.~Huang, P.-Y. Chen, and B.~Li, ``Dba: Distributed backdoor attacks
  against federated learning,'' in \emph{ICLR}, 2019.

\bibitem{wang2020attack}
H.~Wang, K.~Sreenivasan, S.~Rajput, H.~Vishwakarma, S.~Agarwal, J.-y. Sohn,
  K.~Lee, and D.~Papailiopoulos, ``Attack of the tails: Yes, you really can
  backdoor federated learning,'' in \emph{NeurIPS}, 2020.

\bibitem{chen2020backdoor}
C.-L. Chen, L.~Golubchik, and M.~Paolieri, ``Backdoor attacks on federated
  meta-learning,'' \emph{arXiv preprint arXiv:2006.07026}, 2020.

\bibitem{liu2020backdoor}
Y.~Liu, Z.~Yi, and T.~Chen, ``Backdoor attacks and defenses in
  feature-partitioned collaborative learning,'' in \emph{ICML Workshop}, 2020.

\bibitem{liu2022defending}
Y.~Liu, Z.~Yi, Y.~Kang, Y.~He, W.~Liu, T.~Zou, and Q.~Yang, ``Defending label
  inference and backdoor attacks in vertical federated learning,'' in
  \emph{AAAI}, 2022.

\bibitem{sun2019can}
Z.~Sun, P.~Kairouz, A.~T. Suresh, and H.~B. McMahan, ``Can you really backdoor
  federated learning?'' in \emph{NeurIPS Workshop}, 2019.

\bibitem{safa2021defending}
M.~S. Ozdayi, M.~Kantarcioglu, and Y.~R. Gel, ``Defending against backdoors in
  federated learning with robust learning rate,'' in \emph{AAAI}, 2021.

\bibitem{liu2021privacy}
X.~Liu, H.~Li, G.~Xu, Z.~Chen, X.~Huang, and R.~Lu, ``Privacy-enhanced
  federated learning against poisoning adversaries,'' \emph{IEEE Transactions
  on Information Forensics and Security}, vol.~16, pp. 4574--4588, 2021.

\bibitem{yao2019latent}
Y.~Yao, H.~Li, H.~Zheng, and B.~Y. Zhao, ``Latent backdoor attacks on deep
  neural networks,'' in \emph{CCS}, 2019.

\bibitem{chen2022badpre}
K.~Chen, Y.~Meng, X.~Sun, S.~Guo, T.~Zhang, J.~Li, and C.~Fan, ``Badpre:
  Task-agnostic backdoor attacks to pre-trained nlp foundation models,'' in
  \emph{ICLR}, 2022.

\bibitem{adi2018turning}
Y.~Adi, C.~Baum, M.~Cisse, B.~Pinkas, and J.~Keshet, ``Turning your weakness
  into a strength: Watermarking deep neural networks by backdooring,'' in
  \emph{USENIX Security}, 2018.

\bibitem{li2022defending}
Y.~Li, L.~Zhu, X.~Jia, Y.~Jiang, S.-T. Xia, and X.~Cao, ``Defending against
  model stealing via verifying embedded external features,'' in \emph{AAAI},
  2022.

\bibitem{sommer2020towards}
D.~M. Sommer, L.~Song, S.~Wagh, and P.~Mittal, ``Towards probabilistic
  verification of machine unlearning,'' \emph{arXiv preprint arXiv:2003.04247},
  2020.

\bibitem{shan2020using}
S.~Shan, E.~Wenger, B.~Wang, B.~Li, H.~Zheng, and B.~Y. Zhao, ``Using honeypots
  to catch adversarial attacks on neural networks,'' in \emph{CCS}, 2020.

\bibitem{li2020open}
Y.~Li, Z.~Zhang, J.~Bai, B.~Wu, Y.~Jiang, and S.-T. Xia, ``Open-sourced dataset
  protection via backdoor watermarking,'' in \emph{NeurIPS Workshop}, 2020.

\bibitem{zhao2021deep}
S.~Zhao, X.~Ma, Y.~Wang, J.~Bailey, B.~Li, and Y.-G. Jiang, ``What do deep nets
  learn? class-wise patterns revealed in the input space,'' \emph{arXiv
  preprint arXiv:2101.06898}, 2021.

\bibitem{lin2020you}
Y.-S. Lin, W.-C. Lee, and Z.~B. Celik, ``What do you see? evaluation of
  explainable artificial intelligence (xai) interpretability through neural
  backdoors,'' in \emph{KDD}, 2021.

\bibitem{he2016deep}
K.~He, X.~Zhang, S.~Ren, and J.~Sun, ``Deep residual learning for image
  recognition,'' in \emph{CVPR}, 2016.

\bibitem{CIFAR}
A.~Krizhevsky, ``Learning multiple layers of features from tiny images,'' Tech.
  Rep., 2009.

\bibitem{zhang2021inject}
Z.~Zhang, L.~Lyu, W.~Wang, L.~Sun, and X.~Sun, ``How to inject backdoors with
  better consistency: Logit anchoring on clean data,'' \emph{arXiv preprint
  arXiv:2109.01300}, 2021.

\bibitem{guo2020trojannet}
C.~Guo, R.~Wu, and K.~Q. Weinberger, ``Trojannet: Embedding hidden trojan horse
  models in neural networks,'' \emph{arXiv preprint arXiv:2002.10078}, 2020.

\bibitem{mopuri2018generalizable}
K.~R. Mopuri, A.~Ganeshan, and R.~V. Babu, ``Generalizable data-free objective
  for crafting universal adversarial perturbations,'' \emph{IEEE transactions
  on pattern analysis and machine intelligence}, vol.~41, no.~10, pp.
  2452--2465, 2018.

\bibitem{thys2019fooling}
S.~Thys, W.~Van~Ranst, and T.~Goedem{\'e}, ``Fooling automated surveillance
  cameras: adversarial patches to attack person detection,'' in \emph{CVPR
  Workshop}, 2019.

\bibitem{liu2017neural}
Y.~Liu, Y.~Xie, and A.~Srivastava, ``Neural trojans,'' in \emph{ICCD}, 2017.

\bibitem{doan2019februus}
B.~G. Doan, E.~Abbasnejad, and D.~C. Ranasinghe, ``Februus: Input purification
  defense against trojan attacks on deep neural network systems,'' in
  \emph{ACSAC}, 2020.

\bibitem{udeshi2019model}
S.~Udeshi, S.~Peng, G.~Woo, L.~Loh, L.~Rawshan, and S.~Chattopadhyay, ``Model
  agnostic defence against backdoor attacks in machine learning,'' \emph{arXiv
  preprint arXiv:1908.02203}, 2019.

\bibitem{villarreal2020confoc}
M.~Villarreal-Vasquez and B.~Bhargava, ``Confoc: Content-focus protection
  against trojan attacks on neural networks,'' \emph{arXiv preprint
  arXiv:2007.00711}, 2020.

\bibitem{zeng2020deepsweep}
Y.~Zeng, H.~Qiu, S.~Guo, T.~Zhang, M.~Qiu, and B.~Thuraisingham, ``Deepsweep:
  An evaluation framework for mitigating dnn backdoor attacks using data
  augmentation,'' in \emph{AsiaCCS}, 2021.

\bibitem{liu2018fine}
K.~Liu, B.~Dolan-Gavitt, and S.~Garg, ``Fine-pruning: Defending against
  backdooring attacks on deep neural networks,'' in \emph{RAID}, 2018.

\bibitem{zhao2020bridging}
P.~Zhao, P.-Y. Chen, P.~Das, K.~N. Ramamurthy, and X.~Lin, ``Bridging mode
  connectivity in loss landscapes and adversarial robustness,'' in \emph{ICLR},
  2020.

\bibitem{yoshida2020disabling}
K.~Yoshida and T.~Fujino, ``Disabling backdoor and identifying poison data by
  using knowledge distillation in backdoor attacks on deep neural networks,''
  in \emph{CCS Workshop}, 2020.

\bibitem{li2021neural}
Y.~Li, X.~Lyu, N.~Koren, L.~Lyu, B.~Li, and X.~Ma, ``Neural attention
  distillation: Erasing backdoor triggers from deep neural networks,'' in
  \emph{ICLR}, 2021.

\bibitem{wu2021adversarial}
D.~Wu and Y.~Wang, ``Adversarial neuron pruning purifies backdoored deep
  models,'' in \emph{NeurIPS}, 2021.

\bibitem{zeng2022adversarial}
Y.~Zeng, S.~Chen, W.~P. Z.~Morley~Mao, M.~Jin, and R.~Jia, ``Adversarial
  unlearning of backdoors via implicit hypergradient,'' in \emph{ICLR}, 2022.

\bibitem{qiao2019defending}
X.~Qiao, Y.~Yang, and H.~Li, ``Defending neural backdoors via generative
  distribution modeling,'' in \emph{NeurIPS}, 2019.

\bibitem{zhu2020gangsweep}
L.~Zhu, R.~Ning, C.~Wang, C.~Xin, and H.~Wu, ``Gangsweep: Sweep out neural
  backdoors by gan,'' in \emph{ACM MM}, 2020.

\bibitem{cheng2019defending}
H.~Cheng, K.~Xu, S.~Liu, P.-Y. Chen, P.~Zhao, and X.~Lin, ``Defending against
  backdoor attack on deep neural networks,'' in \emph{KDD Workshop}, 2019.

\bibitem{aiken2020neural}
W.~Aiken, H.~Kim, and S.~Woo, ``Neural network laundering: Removing black-box
  backdoor watermarks from deep neural networks,'' \emph{arXiv preprint
  arXiv:2004.11368}, 2020.

\bibitem{harikumar2020scalable}
H.~Harikumar, V.~Le, S.~Rana, S.~Bhattacharya, S.~Gupta, and S.~Venkatesh,
  ``Scalable backdoor detection in neural networks,'' \emph{arXiv preprint
  arXiv:2006.05646}, 2020.

\bibitem{xiang2020detection}
Z.~Xiang, D.~J. Miller, and G.~Kesidis, ``Detection of backdoors in trained
  classifiers without access to the training set,'' \emph{IEEE Transactions on
  Neural Networks and Learning Systems}, 2020.

\bibitem{guo2021aeva}
J.~Guo, A.~Li, and C.~Liu, ``Aeva: Black-box backdoor detection using
  adversarial extreme value analysis,'' in \emph{ICLR}, 2022.

\bibitem{hu2022trigger}
X.~Hu, X.~Lin, M.~Cogswell, Y.~Yao, S.~Jha, and C.~Chen, ``Trigger hunting with
  a topological prior for trojan detection,'' in \emph{ICLR}, 2022.

\bibitem{xu2019detecting}
X.~Xu, Q.~Wang, H.~Li, N.~Borisov, C.~A. Gunter, and B.~Li, ``Detecting ai
  trojans using meta neural analysis,'' in \emph{IEEE S\&P}, 2021.

\bibitem{huangone20}
S.~Huang, W.~Peng, Z.~Jia, and Z.~Tu, ``One-pixel signature: Characterizing cnn
  models for backdoor detection,'' in \emph{ECCV}, 2020.

\bibitem{wang2020practical}
R.~Wang, G.~Zhang, S.~Liu, P.-Y. Chen, J.~Xiong, and M.~Wang, ``Practical
  detection of trojan neural networks: Data-limited and data-free cases,'' in
  \emph{ECCV}, 2020.

\bibitem{zheng2021topological}
S.~Zheng, Y.~Zhang, H.~Wagner, M.~Goswami, and C.~Chen, ``Topological detection
  of trojaned neural networks,'' in \emph{NeurIPS}, 2021.

\bibitem{xiang2022post}
Z.~Xiang, D.~J. Miller, and G.~Kesidis, ``Post-training detection of backdoor
  attacks for two-class and multi-attack scenarios,'' in \emph{ICLR}, 2022.

\bibitem{du2020}
M.~Du, R.~Jia, and D.~Song, ``Robust anomaly detection and backdoor attack
  detection via differential privacy,'' in \emph{ICLR}, 2020.

\bibitem{hong2020effectiveness}
S.~Hong, V.~Chandrasekaran, Y.~Kaya, T.~Dumitra{\c{s}}, and N.~Papernot, ``On
  the effectiveness of mitigating data poisoning attacks with gradient
  shaping,'' \emph{arXiv preprint arXiv:2002.11497}, 2020.

\bibitem{borgnia2021strong}
E.~Borgnia, V.~Cherepanova, L.~Fowl, A.~Ghiasi, J.~Geiping, M.~Goldblum,
  T.~Goldstein, and A.~Gupta, ``Strong data augmentation sanitizes poisoning
  and backdoor attacks without an accuracy tradeoff,'' in \emph{ICASSP}, 2021.

\bibitem{liu2021removing}
X.~Liu, F.~Li, B.~Wen, and Q.~Li, ``Removing backdoor-based watermarks in
  neural networks with limited data,'' in \emph{ICPR}, 2021.

\bibitem{huang2022backdoor}
K.~Huang, Y.~Li, B.~Wu, Z.~Qin, and K.~Ren, ``Backdoor defense via decoupling
  the training process,'' in \emph{ICLR}, 2022.

\bibitem{tran2018spectral}
B.~Tran, J.~Li, and A.~Madry, ``Spectral signatures in backdoor attacks,'' in
  \emph{NeurIPS}, 2018.

\bibitem{chen2019detecting}
B.~Chen, W.~Carvalho, N.~Baracaldo, H.~Ludwig, B.~Edwards, T.~Lee, I.~Molloy,
  and B.~Srivastava, ``Detecting backdoor attacks on deep neural networks by
  activation clustering,'' in \emph{AAAI Workshop}, 2019.

\bibitem{tang2019demon}
D.~Tang, X.~Wang, H.~Tang, and K.~Zhang, ``Demon in the variant: Statistical
  analysis of dnns for robust backdoor contamination detection,'' in
  \emph{USENIX Security}, 2021.

\bibitem{soremekun2020exposing}
E.~Soremekun, S.~Udeshi, S.~Chattopadhyay, and A.~Zeller, ``Exposing backdoors
  in robust machine learning models,'' \emph{arXiv preprint arXiv:2003.00865},
  2020.

\bibitem{chan2019poison}
A.~Chan and Y.-S. Ong, ``Poison as a cure: Detecting \& neutralizing
  variable-sized backdoor attacks in deep neural networks,'' \emph{arXiv
  preprint arXiv:1911.08040}, 2019.

\bibitem{hayase2021spectre}
J.~Hayase and W.~Kong, ``Spectre: Defending against backdoor attacks using
  robust covariance estimation,'' in \emph{ICML}, 2021.

\bibitem{wang2021unified}
T.~Wang, Y.~Zeng, M.~Jin, and R.~Jia, ``A unified framework for task-driven
  data quality management,'' \emph{arXiv preprint arXiv:2106.05484}, 2021.

\bibitem{zeng2021rethinking}
Y.~Zeng, W.~Park, Z.~M. Mao, and R.~Jia, ``Rethinking the backdoor attacks'
  triggers: A frequency perspective,'' in \emph{ICCV}, 2021.

\bibitem{gao2019strip}
Y.~Gao, C.~Xu, D.~Wang, S.~Chen, D.~C. Ranasinghe, and S.~Nepal, ``Strip: A
  defence against trojan attacks on deep neural networks,'' in \emph{ACSAC},
  2019.

\bibitem{subedar2019deep}
M.~Subedar, N.~Ahuja, R.~Krishnan, I.~J. Ndiour, and O.~Tickoo, ``Deep
  probabilistic models to detect data poisoning attacks,'' in \emph{NeurIPS
  Workshop}, 2019.

\bibitem{jin2020unified}
K.~Jin, T.~Zhang, C.~Shen, Y.~Chen, M.~Fan, C.~Lin, and T.~Liu, ``A unified
  framework for analyzing and detecting malicious examples of dnn models,''
  \emph{arXiv preprint arXiv:2006.14871}, 2020.

\bibitem{javaheripi2020cleann}
M.~Javaheripi, M.~Samragh, G.~Fields, T.~Javidi, and F.~Koushanfar, ``Cleann:
  Accelerated trojan shield for embedded neural networks,'' in \emph{ICCAD},
  2020.

\bibitem{dong2019efficient}
Y.~Dong, H.~Su, B.~Wu, Z.~Li, W.~Liu, T.~Zhang, and J.~Zhu, ``Efficient
  decision-based black-box adversarial attacks on face recognition,'' in
  \emph{CVPR}, 2019.

\bibitem{chen2020boosting}
W.~Chen, Z.~Zhang, X.~Hu, and B.~Wu, ``Boosting decision-based black-box
  adversarial attacks with random sign flip,'' in \emph{ECCV}, 2020.

\bibitem{tramer2020adaptive}
F.~Tramer, N.~Carlini, W.~Brendel, and A.~Madry, ``On adaptive attacks to
  adversarial example defenses,'' in \emph{NeurIPS}, 2020.

\bibitem{weng2020trade}
C.-H. Weng, Y.-T. Lee, and S.-H.~B. Wu, ``On the trade-off between adversarial
  and backdoor robustness,'' in \emph{NeurIPS}, 2020.

\bibitem{cohen2019certified}
J.~M. Cohen, E.~Rosenfeld, and J.~Z. Kolter, ``Certified adversarial robustness
  via randomized smoothing,'' in \emph{ICML}, 2019.

\bibitem{selvaraju2017grad}
R.~R. Selvaraju, M.~Cogswell, A.~Das, R.~Vedantam, D.~Parikh, and D.~Batra,
  ``Grad-cam: Visual explanations from deep networks via gradient-based
  localization,'' in \emph{ICCV}, 2017.

\bibitem{kirkpatrick2017overcoming}
J.~Kirkpatrick, R.~Pascanu, N.~Rabinowitz, J.~Veness, G.~Desjardins, A.~A.
  Rusu, K.~Milan, J.~Quan, T.~Ramalho, A.~Grabska-Barwinska \emph{et~al.},
  ``Overcoming catastrophic forgetting in neural networks,'' \emph{Proceedings
  of the national academy of sciences}, vol. 114, no.~13, pp. 3521--3526, 2017.

\bibitem{garipov2018loss}
T.~Garipov, P.~Izmailov, D.~Podoprikhin, D.~P. Vetrov, and A.~G. Wilson, ``Loss
  surfaces, mode connectivity, and fast ensembling of dnns,'' in
  \emph{NeurIPS}, 2018.

\bibitem{hinton2015distilling}
G.~Hinton, O.~Vinyals, and J.~Dean, ``Distilling the knowledge in a neural
  network,'' in \emph{NeurIPS Workshop}, 2014.

\bibitem{auer2002finite}
P.~Auer, N.~Cesa-Bianchi, and P.~Fischer, ``Finite-time analysis of the
  multiarmed bandit problem,'' \emph{Machine learning}, vol.~47, no.~2, pp.
  235--256, 2002.

\bibitem{devries2017cutout}
T.~DeVries and G.~W. Taylor, ``Improved regularization of convolutional neural
  networks with cutout,'' \emph{arXiv preprint arXiv:1708.04552}, 2017.

\bibitem{feinman2017detecting}
R.~Feinman, R.~R. Curtin, S.~Shintre, and A.~B. Gardner, ``Detecting
  adversarial samples from artifacts,'' \emph{arXiv preprint arXiv:1703.00410},
  2017.

\bibitem{ma2018characterizing}
X.~Ma, B.~Li, Y.~Wang, S.~M. Erfani, S.~Wijewickrema, G.~Schoenebeck, D.~Song,
  M.~E. Houle, and J.~Bailey, ``Characterizing adversarial subspaces using
  local intrinsic dimensionality,'' in \emph{ICLR}, 2018.

\bibitem{wang2019adversarial}
J.~Wang, G.~Dong, J.~Sun, X.~Wang, and P.~Zhang, ``Adversarial sample detection
  for deep neural network through model mutation testing,'' in \emph{ICSE},
  2019.

\bibitem{MNIST}
Y.~{Lecun}, L.~{Bottou}, Y.~{Bengio}, and P.~{Haffner}, ``Gradient-based
  learning applied to document recognition,'' \emph{Proceedings of the IEEE},
  vol.~86, no.~11, pp. 2278--2324, 1998.

\bibitem{umer2020targeted}
M.~Umer, G.~Dawson, and R.~Polikar, ``Targeted forgetting and false memory
  formation in continual learners through adversarial backdoor attacks,''
  \emph{arXiv preprint arXiv:2002.07111}, 2020.

\bibitem{gao2020analyzing}
Y.~Gao, H.~Rosenberg, K.~Fawaz, S.~Jha, and J.~Hsu, ``Analyzing accuracy loss
  in randomized smoothing defenses,'' \emph{arXiv preprint arXiv:2003.01595},
  2020.

\bibitem{fashionMNIST}
H.~Xiao, K.~Rasul, and R.~Vollgraf, ``Fashion-mnist: a novel image dataset for
  benchmarking machine learning algorithms,'' \emph{arXiv preprint
  arXiv:1708.07747}, 2017.

\bibitem{veldanda2020nnoculation}
A.~K. Veldanda, K.~Liu, B.~Tan, P.~Krishnamurthy, F.~Khorrami, R.~Karri,
  B.~Dolan-Gavitt, and S.~Garg, ``Nnoculation: Broad spectrum and targeted
  treatment of backdoored dnns,'' \emph{arXiv preprint arXiv:2002.08313}, 2020.

\bibitem{SVHN}
Y.~Netzer, T.~Wang, A.~Coates, A.~Bissacco, B.~Wu, and A.~Y. Ng, ``Reading
  digits in natural images with unsupervised feature learning,'' in
  \emph{NeurIPS Workshop}, 2011.

\bibitem{ImageNet}
J.~Deng, W.~Dong, R.~Socher, L.-J. Li, K.~Li, and L.~Fei-Fei, ``Imagenet: A
  large-scale hierarchical image database,'' in \emph{CVPR}, 2009.

\bibitem{GTSRB}
J.~Stallkamp, M.~Schlipsing, J.~Salmen, and C.~Igel, ``Man vs. computer:
  Benchmarking machine learning algorithms for traffic sign recognition,''
  \emph{Neural networks}, vol.~32, pp. 323--332, 2012.

\bibitem{USTS}
A.~Mogelmose, M.~M. Trivedi, and T.~B. Moeslund, ``Vision-based traffic sign
  detection and analysis for intelligent driver assistance systems:
  Perspectives and survey,'' \emph{IEEE Transactions on Intelligent
  Transportation Systems}, vol.~13, no.~4, pp. 1484--1497, 2012.

\bibitem{YouTubeFace}
L.~Wolf, T.~Hassner, and I.~Maoz, ``Face recognition in unconstrained videos
  with matched background similarity,'' in \emph{CVPR}, 2011.

\bibitem{PubFig}
N.~Kumar, A.~C. Berg, P.~N. Belhumeur, and S.~K. Nayar, ``Attribute and simile
  classifiers for face verification,'' in \emph{ICCV}, 2009.

\bibitem{BMVC}
O.~M. Parkhi, A.~Vedaldi, and A.~Zisserman, ``Deep face recognition,'' in
  \emph{BMVC}, 2015.

\bibitem{vggface2}
Q.~Cao, L.~Shen, W.~Xie, O.~M. Parkhi, and A.~Zisserman, ``Vggface2: A dataset
  for recognising faces across pose and age,'' in \emph{IEEE FGR}, 2018.

\bibitem{LFW}
G.~B. Huang, M.~Ramesh, T.~Berg, and E.~Learned-Miller, ``Labeled faces in the
  wild: A database for studying face recognition in unconstrained
  environments,'' University of Massachusetts, Amherst, Tech. Rep. 07-49,
  October 2007.

\bibitem{Tan2020Bypassing}
T.~J.~L. Tan and R.~Shokri, ``Bypassing backdoor detection algorithms in deep
  learning,'' in \emph{EuroS\&P}, 2020.

\bibitem{Aniruddha2020Hidden}
A.~Saha, A.~Subramanya, and H.~Pirsiavash, ``Hidden trigger backdoor attacks,''
  in \emph{AAAI}, 2020.

\bibitem{Rosenfeld2020Certified}
E.~Rosenfeld, E.~Winston, P.~Ravikumar, and J.~Z. Kolter, ``Certified
  robustness to label-flipping attacks via randomized smoothing,'' in
  \emph{ICML}, 2020.

\bibitem{jia2021intrinsic}
J.~Jia, X.~Cao, and N.~Z. Gong, ``Intrinsic certified robustness of bagging
  against data poisoning attacks,'' in \emph{AAAI}, 2021.

\bibitem{levine2021deep}
A.~Levine and S.~Feizi, ``Deep partition aggregation: Provable defenses against
  general poisoning attacks,'' in \emph{ICLR}, 2021.

\bibitem{jia2022certified}
J.~Jia, X.~Cao, and N.~Z. Gong, ``Certified robustness of nearest neighbors
  against data poisoning attacks,'' in \emph{AAAI}, 2022.

\bibitem{breiman1996bagging}
L.~Breiman, ``Bagging predictors,'' \emph{Machine learning}, vol.~24, no.~2,
  pp. 123--140, 1996.

\bibitem{sugiyama2015introduction}
M.~Sugiyama, \emph{Introduction to statistical machine learning}.\hskip 1em
  plus 0.5em minus 0.4em\relax Morgan Kaufmann, 2015.

\end{thebibliography}

\begin{IEEEbiography}[{\includegraphics[width=1in,height=1.25in,clip,keepaspectratio]{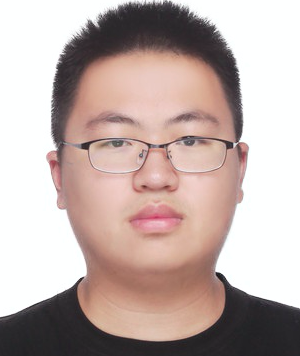}}]{Yiming Li} is currently a Ph.D. candidate from Tsinghua-Berkeley Shenzhen Institute, Tsinghua Shenzhen International Graduate School, Tsinghua University. Before that, he received his B.S. degree in Mathematics and Applied Mathematics from Ningbo University in 2018. His research interests are in the domain of AI security, especially backdoor learning, adversarial learning, and data privacy. His researches have been published in multiple top-tier conferences and journals, such as ICCV, ECCV, ICLR, AAAI, PR Journal, and IEEE IoT Journal. He served as the senior program committee member of AAAI 2022 and the reviewer of IEEE TDSC, IEEE TCSVT, IEEE TII, Neurocomputing, etc.
\end{IEEEbiography}

\begin{IEEEbiography}[{\includegraphics[width=1in,height=1.25in,clip,keepaspectratio]{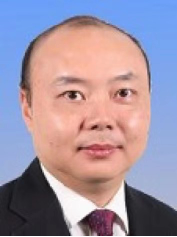}}]{Dr. Yong Jiang} received his M.S. and Ph.D. degrees in computer science from Tsinghua University, China, in 1998 and 2002, respectively. Since 2002, he has been with the Tsinghua Shenzhen International Graduate School of Tsinghua University, Guangdong, China, where he is currently a full professor. His research interests include computer vision, machine learning, Internet architecture and its protocols, IP routing technology, etc. He has received several best paper awards (e.g., IWQoS 2018) and his researches have been published in multiple top-tier journals and conferences, including IEEE ToC, IEEE TMM, IEEE TSP, CVPR, ICLR, ECCV, etc.
\end{IEEEbiography}

\begin{IEEEbiography}[{\includegraphics[width=1in,height=1.25in,clip,keepaspectratio]{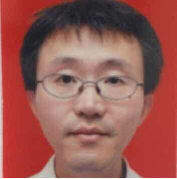}}]{Dr. Zhifeng Li} is currently a top-tier principal researcher at Tencent Data Platform. He received the Ph.D. degree from the Chinese University of Hong Kong in 2006. After that, He was a postdoctoral fellow at the Chinese University of Hong Kong and Michigan State University for several years. Before joining Tencent, he was a full professor with the Shenzhen Institutes of Advanced Technology, Chinese Academy of Sciences. His research interests include deep learning, computer vision and pattern recognition, and face detection and recognition. He is currently serving on the Editorial Boards of Pattern Recognition, IEEE Transactions on Circuits and Systems for Video Technology, and Neurocomputing. He is a fellow of the British Computer Society (FBCS).
\end{IEEEbiography}

\begin{IEEEbiography}[{\includegraphics[width=1in,height=1.25in,clip,keepaspectratio]{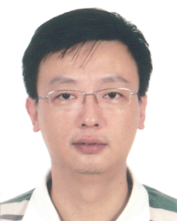}}]{Dr. Shu-Tao Xia} received the B.S. degree in mathematics and the Ph.D. degree in applied mathematics from Nankai University, Tianjin, China, in 1992 and 1997, respectively. Since January 2004, he has been with the Tsinghua Shenzhen International Graduate School of Tsinghua University, Guangdong, China, where he is currently a full professor. From March 1997 to April 1999, he was with the research group of information theory, Department of Mathematics, Nankai University, Tianjin, China. From September 1997 to March 1998 and from August to September 1998, he visited the Department of Information Engineering, The Chinese University of Hong Kong, Hong Kong. His current research interests include coding and information theory, machine learning, and deep learning. His researches have been published in multiple top-tier journals and conferences, including IEEE TIP, IEEE TNNLS, CVPR, ICCV, ECCV, ICLR, etc.
\end{IEEEbiography}

\end{document}